%% file: main_v3.tex
\documentclass[a4paper,fleqn,usenatbib,useAMS]{mnras}

\usepackage{graphicx}	
\usepackage{amsmath}	
\usepackage{amssymb}	
\usepackage{multicol}        
\usepackage{bm}		
\usepackage{pdflscape}	
\usepackage{natbib}
\usepackage{times}
\usepackage{calc}
\usepackage{rotating}
\usepackage{lscape}
\bibpunct{(}{)}{;}{a}{}{,} 
\usepackage{longtable}
\usepackage[referable]{threeparttablex}
\usepackage{threeparttable}
\usepackage{enumitem} 
\usepackage{hyperref}
\usepackage{pifont}
\usepackage{tikz}
\usepackage{xcolor}
\usepackage{ulem}

\def\oversim#1#2{\lower0.5pt\vbox{\baselineskip0pt \lineskip-0.5pt
     \ialign{$\mathsurround0pt #1\hfil##\hfil$\crcr#2\crcr\sim\crcr}}}

\newcommand{\useaa}{} 
\newcommand{\forb}[2]{[\ion{#1}{#2}]}
\newcommand{\forbl}[3]{\forb{#1}{#2} $\lambda$#3\useaa}
\newcommand{\forbr}[4]{\forb{#1}{#2} $\lambda\lambda$#3/#4\useaa}
\newcommand{\forbs}[4]{\forb{#1}{#2} $\lambda\lambda$#3, #4\useaa}

\newcommand{\perm}[2]{\ion{#1}{#2}}
\newcommand{\perml}[3]{\perm{#1}{#2} $\lambda$#3\useaa}
\newcommand{\perms}[4]{\perm{#1}{#2} $\lambda\lambda$#3, #4\useaa}

\usepackage[T1]{fontenc}
\usepackage{ae,aecompl}

\usepackage{mathptmx}

\def\apj {{ApJ}}

\def\aap {{A\&A}}
\def\mnras {{MNRAS}}

\def\apjl {{ApJ}}

\def\apjs {{ApJs}}

\title[Properties of PC\,22]{Catching a grown-up starfish planetary nebula -- II. Plasma analysis and central star properties of PC\,22 }

\author[L. Sabin et al.] 
{L. Sabin$^{1}$\thanks{E-mail:lsabin@astro.unam.mx (LS)}, V. G\'omez-Llanos$^{1}$, C. Morisset$^{1}$,V.~M.~A. G\'{o}mez-Gonz\'{a}lez$^{2,4}$, M.A. Guerrero$^{3}$ 
\newauthor H. Todt$^{4}$ and X. Fang$^{5,6}$ \\  
$^{1}$Instituto de Astronomía, Universidad Nacional Autónoma de México, AP 106,  Ensenada 22800, BC, Mexico\\
$^{2}$Instituto de Radioastronom\'ia y Astrofísica (IRyA), UNAM Campus Morelia, Apartado postal 3-72, 58090 Morelia, Michoac\'an, Mexico \\
$^{3}$Instituto de Astrof\'{\i}sica de Andaluc\'{\i}a, IAA-CSIC, C/ Glorieta de la Astronom\'{\i}a s/n, 18008 Granada, Spain\\
$^{4}$Institute for Physics and Astronomy, Universit$\ddot{a}$t Potsdam, Karl-Liebknecht-Str. 24/25, D-14476 Potsdam, Germany \\
$^{5}$Key Laboratory of Optical Astronomy, National Astronomical Observatories, Chinese Academy of Sciences (NAOC), Beijing, China \\
$^{6}$Department of Physics \& Laboratory for Space Research, Faculty of Science, University of Hong Kong, Hong Kong, China \\}

\date{Accepted 2021 December 13. Received 2021 December 9; in original form 2021 June 2}

\pubyear{2021}

\begin{document}
\label{firstpage}
\pagerange{\pageref{firstpage}--\pageref{lastpage}}
\maketitle


\begin{abstract}

After performing the morpho-kinematic analysis of the planetary nebula (PN) PC\,22, we now present its nebular and stellar analysis. The plasma investigation relies on the novel use of a Monte Carlo analysis associated to the {\sc PyNeb} code for the
uncertainty propagation. The innermost region of the nebula shows electronic temperatures $T_{\rm e}$ $\approx$ 10,800~K using \forb{N}{ii} and $\approx$ 13,000 K using \forb{O}{iii} and electronic densities $n_{\rm e}$ $\approx$ 600~cm$^{-3}$. We also used for the first time a Machine Learning Algorithm to calculate Ionisation Correction Factors (ICFs) specifically adapted to PC\,22. This has allowed us to have pioneer ICFs for (S$^{+}$ + S$^{++}$)/O$^{++}$, Cl$^{++}$/O$^{++}$, and Ar$^{3+}$+Ar$^{4+}$, as well as a possible new determination for the total abundance of neon. The study of the stellar spectrum revealed the presence of broad emission lines consistent with a Wolf-Rayet-type [WR] classification and more precisely a [WO1]-subtype based on different qualitative and quantitative criteria. This classification is also coherent with the high stellar temperature derived from the reproduction of the ionization state of the gas with the Mexican Million Models database (3MdB) and the best fit model obtained with the NLTE model atmosphere code PoWR. PC\,22 is therefore a new addition to the [WO1]-subtype PNe.
  
\end{abstract}


\section{Introduction}

In a previous 
work, \citet[hereafter Paper I]{Sabin2017} performed the morpho-kinematics analysis of the planetary nebula (PN) PC\,22 and classified it as an evolved starfish with fast outflows. 
The deep imaging used for the morphological analysis also revealed an interesting spatial distribution for the \forbl{O}{iii}{5007}~nebular emission. The latter appeared as the dominant emission line not only in the central part of the PN, but also in its outermost regions, i.e. the faint lobes, therefore probing the high excitation state of PC\,22. This PN could therefore be compared to other highly ionised objects such as NGC 6309 \citep{Rubio2015}, NGC 6905 \citep{Cuesta1993,Gomez2022} and NGC 6058  \citep{Guillen2013} for example.

In this second part of our investigation of PC\,22 we use optical long slit spectroscopic observations of its nebular and stellar components to determine the chemical abundances of the PN and to characterise its central star (CSPN). 
Amongst the highlights of this paper is also the first time application of the Mexican Million Models database (3MdB, \citealt{Morisset2015}) coupled with a Machine Learning Algorithm to study the PN plasma (i.e. electronic temperatures and densities, ionic and elemental abundances). With this information it will therefore be possible to obtain a comprehensive analysis of PC\,22, but also to perform some comparative studies with other high excitation PNe. \\
\indent
The article is organized as follows. The observations are presented in section \S\ref{sec_obs}. The results of the nebular and stellar analysis are shown in \S\ref{Neb} and \S\ref{CS_analysis}, respectively. Finally our discussion and concluding remarks are presented in section \S\ref{sec_disc} and \S\ref{sec_con}, respectively.

\section[]{Observations}\label{sec_obs}

Long-slit intermediate resolution optical spectroscopic observations of PC\,22 were performed on July 19, 2015 with the Alhambra Faint Object Spectrograph and Camera (ALFOSC) mounted on the 2.5m Nordic Optical Telescope (NOT) at the Roque de los Muchachos Observatory (ORM, La Palma, Spain).
The detector was an E2V 42-40 2k$\times$2k CCD with pixel size 15\,$\mu$m, providing a plate scale of 0$\farcs$2138 pixel$^{-1}$ and a field of view (FoV) of 6$\farcm$3$\times$6$\farcm$3. 
The grism \#7, providing a dispersion of 1.7\AA\, pixel$^{-1}$, was used in conjunction with the grism \#14, which provides a dispersion of 1.6\AA\, pixel$^{-1}$. The use of both grisms allowed a total coverage from 3200\AA\, to 7110\AA.
The slit length matches the FoV and a slit width of 0$\farcs$55 was selected. 
This slit width provides a spectral resolution of $\Delta\lambda$= 4.1\AA\, for the grism \#7 and $\Delta\lambda$= 3.8\AA\, for the grism \#14. 

\begin{figure}
\begin{center}
\includegraphics[width=\linewidth]{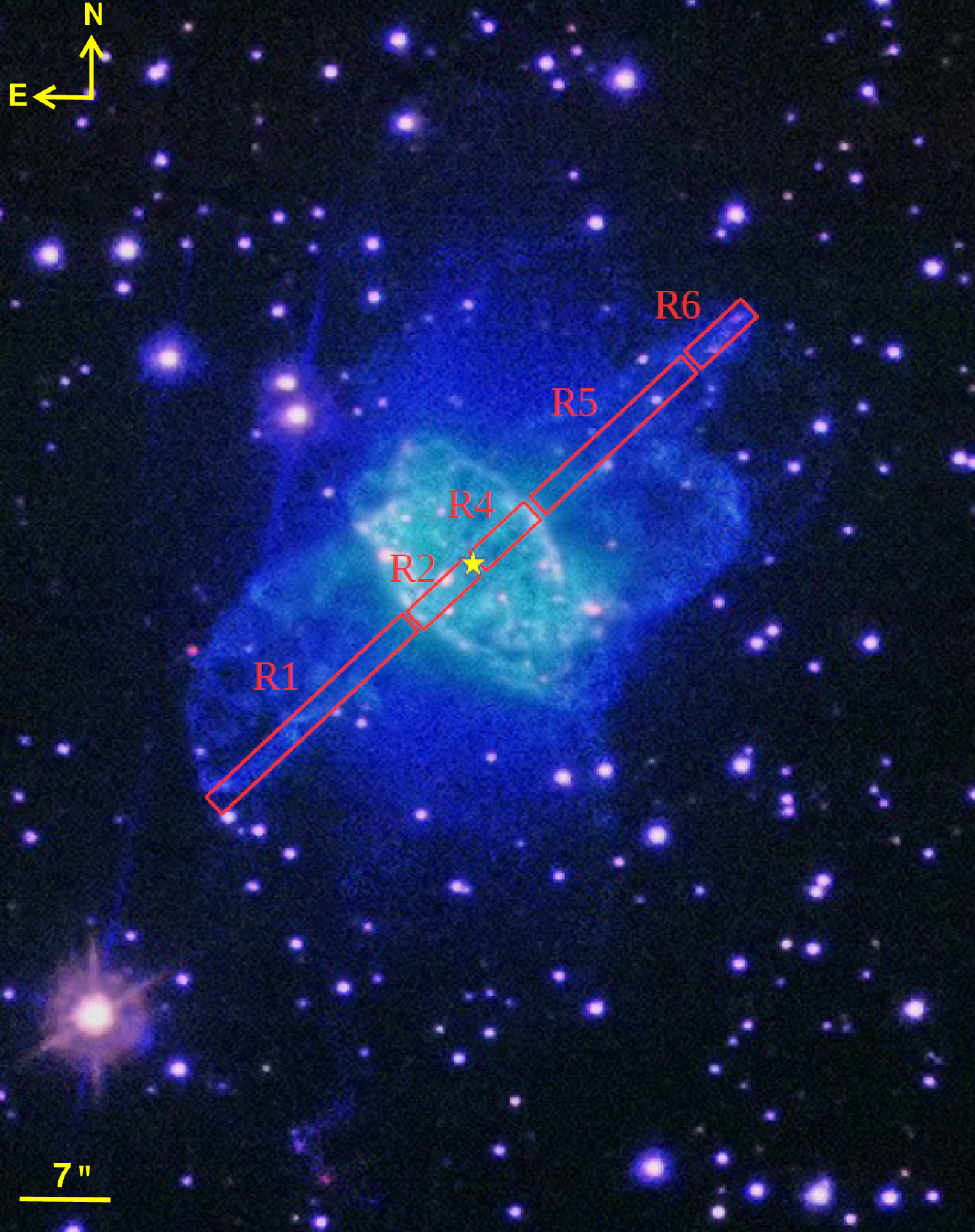}
\caption{
Colour-composite Nordic Optical Telescope picture of PC 22 as presented in Paper I (R=H$\alpha$, G=[NII], B=[OIII]). The slit used for the acquisition of intermediate-resolution spectroscopy is represented by the red segments and is divided in 6 regions of interest (see Table \ref{data}), including the central star marked as a yellow star (see Tables \ref{tab:elines}). The slit characteristics are presented in \S 2. 
North is up and East is left}
\label{ImageNOT}
\end{center}
\end{figure}

The slit was placed along the bright internal elliptical structure's minor axis of PC\,22 at position angle (PA) of 133$^{\circ}$ (see Fig.~\ref{ImageNOT}). 
This slit allows us to analyse the external outflows and internal bright elliptical structure of PC\,22, as well as its CSPN.
The exposure time was set to 2$\times$1500s with grism \#7 and 2$\times$600s with grism \#14.

The data reduction was performed following standard procedures with {\sc IRAF} and includes bias subtraction, flat-fielding, wavelength calibration with a helium-neon arc lamp and finally flux calibration using the standard star Feige~110.

\begin{table*} 
\caption{Observed and dereddened fluxes derived for the different nebular zones shown in Fig.\ref{ImageNOT}. All values are with respect to H$\beta$=100.  ($^{*}$) In the case of Regions 1 and 5 the errors on the derivation of c(H$\beta$) are $\geq$ 50\%.} 
\label{data}
\scalebox{0.75}{
\begin{tabular}{@{\extracolsep{4pt}}lrcccccccc} \hline \hline 
\multicolumn{1}{|l|}{} &\multicolumn{1}{|r|}{} & \multicolumn{2}{c}{{\bf Region 1}} & \multicolumn{2}{c|}{{\bf Region 2}} &  \multicolumn{2}{c|}{{\bf Region 4}}& \multicolumn{2}{c}{{\bf Region 5}}\\  
\cline{3-4} \cline{5-6} \cline{7-8} \cline{9-10}  
\multicolumn{1}{|l|}{Ion}&  \multicolumn{1}{|r|}{$\lambda$ (\AA)}  & \multicolumn{1}{|c|}{F/H$\beta$} & \multicolumn{1}{|c|}{(I/I$\beta)$}   &     \multicolumn{1}{|c|}{F/H$\beta$}  &  \multicolumn{1}{|c|}{(I/I$\beta)$}  & \multicolumn{1}{|c|}{F/H$\beta$} & \multicolumn{1}{|c|}{(I/I$\beta)$}   &    
\multicolumn{1}{|c|}{F/H$\beta$}  &  \multicolumn{1}{|c|}{(I/I$\beta)$}      \\    
\hline
\input{tab1_final3}

\hline                                                                                 
\end{tabular}}                                                                         
\end{table*}                                                                           
                         

\begin{table} 
\centering
\caption{Dereddened fluxes derived for the bright central area. A systematic error of 10\% has been quadratically added to the observation error.} 
\label{centralNeb}
\scalebox{1.0}{
\begin{tabular}{@{\extracolsep{4pt}}lcc} \hline \hline 
\multicolumn{1}{|l|}{} &\multicolumn{1}{|r|}{} & \multicolumn{1}{c}{{\bf Central Area}}\\
 \cline{3-3} 
\multicolumn{1}{|l|}{Ion}&  \multicolumn{1}{|r|}{$\lambda$ (\AA)} & \multicolumn{1}{|c|}{(I/I$\beta)$} \\    
\hline
\input{tab2_final3}

\end{tabular}}                                                                         
\end{table}

%


\section{Nebular analysis}\label{Neb}

We identified and extracted five nebular zones along the minor axis of the internal elliptical structure of PC\,22 using the {\sc IRAF} task {\it apall} (Fig.~\ref{ImageNOT}). Two of these regions are located inside the main elliptical structure (named R2 and R4) and three correspond to the opposite outflows (named R1, R5 and R6). The central star (R3) has also been extracted and its analysis is presented in section \S \ref{CS_analysis}. 

The presence of strong \perml{He}{ii}{4686} emission in the whole nebular spectrum of PC\,22 denotes a highly excited nebula. 
The hydrogen Balmer lines are therefore likely to be affected by the contribution from unresolved \perm{He}{ii} Pickering lines. We calculated an average contamination level of 6\% on the basis of \citet{Brocklehurst1971}'s theoretical Balmer line ratios. 
The corrective procedure, mostly valid for the regions R2 and R4 (see below), does not affect the extinction measurement but can impact the abundances determination (if not performed).


\subsection{The faint outflows}\label{FO}

Table\,\ref{data} indicates that the spatially most extreme regions, corresponding to the South-East and North-West outflows (R1, R5 and R6) present a relatively scarce number of emission lines.
Hence, for the regions R1 and R5 we only detected the Balmer lines H$\alpha$ and H$\beta$, as well as \forbl{N}{ii}{6584} (not in R1), \forbs{O}{iii}{4959}{5007}, and interestingly \perml{He}{ii}{4686}. Due to the faintness of the H$\beta$ line, the error on the extinction determination, using the \citet{Cardelli1989} law with R$_{v}$=3.1\footnote{This law is used throughout the article}, is large ($\geq$ 50\%). The logarithmic extinction c(H$\beta$) values for R1 and R5 show a large variation from  0.6 to 0.1 respectively. \\
In the case of the region R6, the H$\beta$ line is absent and therefore impedes the dereddening of the detected emission lines (i.e. H$\alpha$, \forbs{N}{ii}{6548}{6584} and \forbs{O}{iii}{4959}{5007}). We therefore chose not to include this region in Table \ref{data} although it is worth noting though that if we consider H$\alpha$=100, \forbl{O}{iii}{5007} = 1140$\pm$130.

As inferred from the analysis of the images in Paper I, the \forbl{O}{iii}{5007} line is dominant and we now quantify that it is up to 15.5 times the H$\beta$ value. The outflows also display a high ionization degree with \perml{He}{ii}{4686} to H$\beta$ ratios of 1.2 and 1.6 in R1 and R5, respectively. 
The high value in the latter could be associated to the presence of a denser filamentary area. The tip of the North-West outflow (R6) does not show this emission at all.
It was not possible to derive the electronic temperatures (T$_\mathrm{e}$) and densities (n$_\mathrm{e}$) in these external regions due to the lack of appropriate diagnostic emission lines, nor was it possible to obtain the chemical abundances. 


\subsection{The bright elliptical structure}\label{BES}

Based on Table \ref{data}, most of the emission is seen in the internal regions, namely R2 and R4, which correspond to the bright central regions of PC\,22. The determination of c(H$\beta$) for the two regions is obtained from their respective H$\alpha$/H$\beta$ line ratio and we found an identical value of 0.32.\\
However, a careful analysis of both areas does not indicate substantial differences in terms of the presence of the different lines identified and their fluxes. We therefore decided to combine both R2 and R4 spectra, and extract the emission lines again with {\it apall}. The new results are shown in Table \ref{centralNeb}. \\
The nebular analysis presented in the following was carried out with {\sc PyNeb} version 1.1.15b2. The atomic data used to determine the ionic abundances are listed in Table~\ref{tab:atomicdata}.

\subsubsection{Error propagation using a Monte Carlo analysis}

We quadratically add 10\% to the errors based on the noise close to the lines, to take into account the additional uncertainties on the various calibrations performed and to anticipate the uncertainties on the atomic data used in the processes applied to the line intensities when deriving the abundances. 
To explore the uncertainty propagation from the line intensities to the final abundances, we adopt a Monte Carlo (MC) based approach generating 5,000 artificial observations (above this number the statistics does not significantly change) with intensities following a normal distribution around the observed uncorrected values with a standard deviation corresponding to the uncertainty of each line. The MC is automatically done using the {\tt addMonteCarloObs} method of the {\tt Observations} class in {\sc PyNeb}. Each emission line intensity is now a vector of 5,001 values (the first one being the original observed value). We use the uncorrected values to generate the MC data set, the reddening correction being also a source of uncertainty. 
The pipeline used to determine the electronic temperatures and densities (T$_\mathrm{e}$, n$_\mathrm{e}$), the ionic abundances and the final elemental abundances, described in the following sections, is applied to each one of the 5,001 "observations". The computation is made using the parallel mode of {\sc PyNeb} ({\tt pn.config.use\_multiprocs()}). In the following sections, we will refer to the values of physical parameters obtained using the original observed intensities as the "original" method, while the mean or median values as well as the standard deviations are obtained from the MC distributions.
We explored the effect of the systematic 10\% added to the line intensities by reducing this value to 5\%: all the derived parameters remain the same, the only change being the standard deviation, reduced by 0.03 to 0.06 dex, depending on the emission lines involved in their determination.

\subsubsection{Logarithmic extinction c(H$\beta$)}

To take into account uncertainties due to this reddening correction, we compute the value of c(H$\beta$)$_\alpha$ from only H$\alpha$/H$\beta$, and c(H$\beta$)$_{\alpha\gamma\delta}$ from the mean of the values determined from H$\alpha$/H$\beta$, H$\gamma$/H$\beta$ and H$\delta$/H$\beta$. 
We then determine a coefficient $C$ uniformly randomly distributed between 0 and 1 to compute the final value as c(H$\beta$) = C *  c(H$\beta$)$_{\alpha\gamma\delta}$  + (1-C) * c(H$\beta$)$_\alpha$. This procedure is applied for each realisation of the 5000 MC observations. The resulting mean value in our MC distribution for c(H$\beta$) is 0.35$\pm$ 0.13, corresponding to H$\alpha$/H$\beta$ = 2.95 $\pm$ 0.2. The "original" value of c(H$\beta$) using c(H$\beta$)$_{\alpha\gamma\delta}$ is 0.28.
We note that while the extinction factor R$_{v}$ can be another source of uncertainty, we did not consider it as such in our calculations.

\subsubsection{T$_\mathrm{e}$ and n$_\mathrm{e}$ diagnostics}


\begin{table*} 
\centering
\caption{Atomic data from PyNeb v.1.1.15b2 used to compute the physical properties and the ionic abundances of PC~22.} 
\label{tab:atomicdata}
\scalebox{1.0}{
\begin{tabular}{@{\extracolsep{4pt}}lll} \hline \hline 
\multicolumn{1}{|c|}{ion} &\multicolumn{1}{|c|}{recombination coeffs} & \multicolumn{1}{c}{{collision strengths}}\\
\hline
\perm{H}{i}  & \citet{1995Storey_mnra272}& \\
\perm{He}{i} & \citet{2012Porter_mnra425} \nocite{2013Porter_mnra433} &\\
\perm{He}{ii} & \citet{1995Storey_mnra272}& \\
\forb{N}{ii} & \citet{2004Froese-Fischer_Atom87} &  \citet{2011Tayal_apjs195} \\
\forb{O}{ii} & \citet{1982Zeippen_mnra198}& \citet{2009Kisielius_mnra397}\\
\forb{O}{iii} & \citet{2004Froese-Fischer_Atom87} & \citet{2014Storey_mnra441} \\
\forb{Ne}{iii} & \citet{1997Galavis_aaps123}& \citet{2000McLaughlin_Jour33}\\
\forb{Ne}{iv} & \citet{1984Godefroid_Jour17}& \citet{1981Giles_mnra195}\\
\forb{Ne}{v} & \citet{1997Galavis_aaps123}& \citet{2013Dance_mnra}\\
\forb{S}{ii} & \citet{2019Rynkun_aap623} & \citet{2010Tayal_apjs188}\\
\forb{S}{iii} & \citet{2006Froese-Fischer_Atom92} & \citet{1999Tayal_apj526}\\
\forb{Cl}{iii} & \citet{2019Rynkun_aap623}& \citet{1989Butler_aap208}\\
\forb{Ar}{iv} & \citet{2019Rynkun_aap623} & \citet{1997Ramsbottom_Atom66}\\
\forb{Ar}{v} & \citet{1982Mendoza_mnra199} &  \citet{1995Galavis_aaps111}\\
\hline                                                                                 

\end{tabular}}                                                                         
\end{table*}                                                                           

\begin{figure}
\begin{center}
\includegraphics[scale = 0.65]{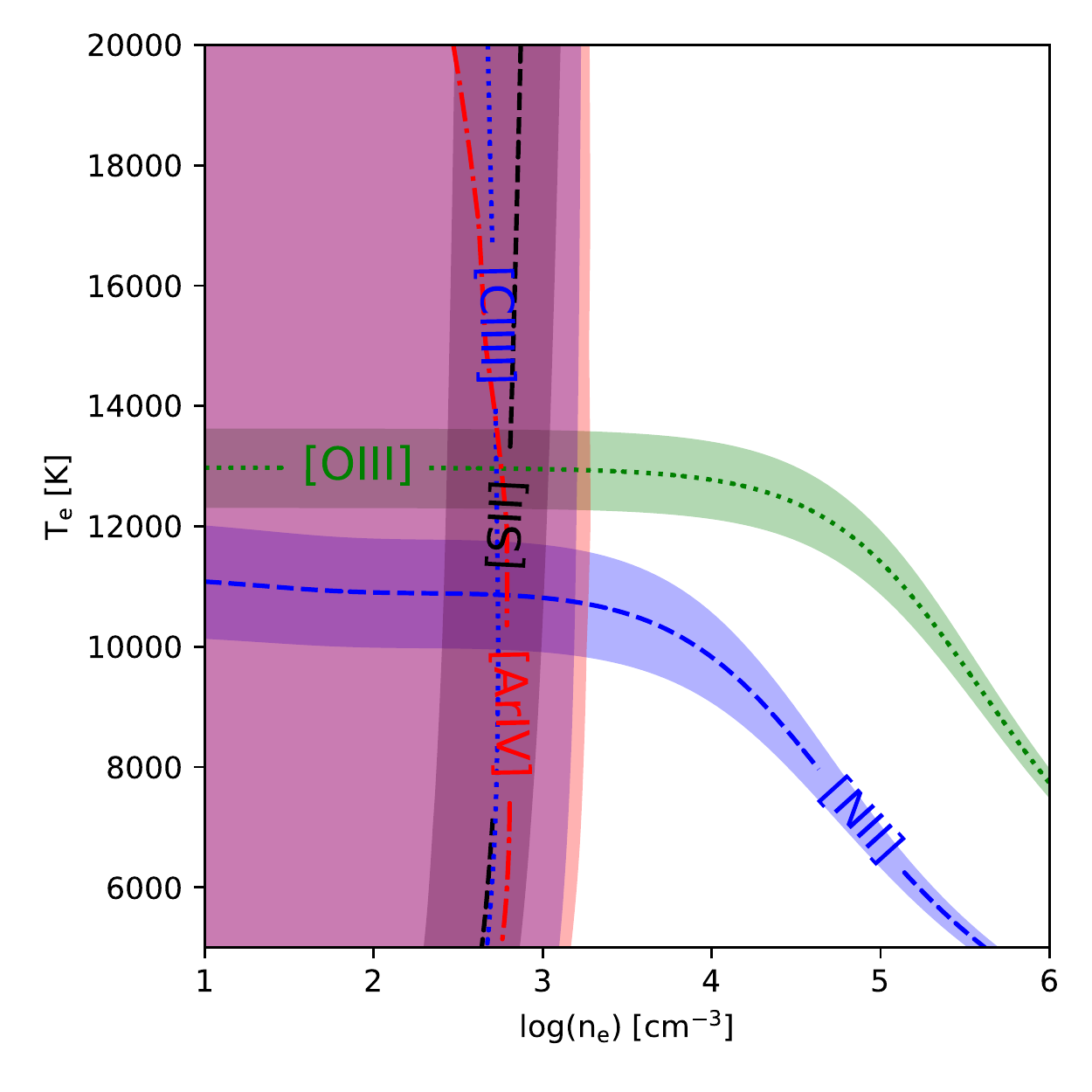}
\caption{Diagnostic diagrams obtained with {\sc PyNeb} for the internal region showing the diagnostics lines used to determine T$_\mathrm{e}$ and n$_\mathrm{e}$ from the "original" values}, based on various ions and the errors on their measurements. 

\label{Cut}
\end{center}
\end{figure}

First we present in Figure \ref{Cut} the classical diagnostic diagram of the "original" values, obtained with the {\tt Diagnostic.plot()} method, showing the intersections of the temperature-sensitive (\,\forbr{O}{iii}{4363}{4959+5007} and \forbr{N}{ii}{5755}{6548+6584}) and density-sensitive line ratios (\forbr{S}{ii}{6731}{6716}, \forbr{Ar}{iv}{4740}{4711} and \forbr{Cl}{iii}{5538}{5518}). We adopted a 2-regions description of the plasma: a low ionization region where the \forb{N}{ii}  and \forb{S}{ii} lines arise, and a high ionization region emitting \forb{O}{iii}, \forb{Cl}{iii} and \forb{Ar}{iv} lines.  We present the results of the electron temperature and density in the upper part of Table~\ref{diagnostics}.
We obtained electronic temperatures of $\approx$10,900 K using \forb{N}{ii} and $\approx$13,000 K using \forb{O}{iii}. As the \forb{O}{iii} temperatures obtained with \forb{Cl}{iii} and \forb{Ar}{iv} are roughly identical, we adopted the mean value for the high ionization region. From Figure~\ref{Cut} (or using {\tt getCrossTemDen} {\sc PyNeb} method) we can determine the electron density for the 3 diagnostics. They all indicate the same order of magnitude between 500 and 630 cm$^{-3}$.

It is important to notice here that the high ionization density diagnostics (\forb{Cl}{iii} and \forb{Ar}{iv}) are both sensitive in a high density range\footnote{The density diagnostic line ratios are sensitive to density variations in a range between the so-called low- and high-density limits.  Outside this range, the line ratios become insensitive to density variations.  In this work the have limited the use of density-sensitive line ratios to values corresponding to 10\% above the low-density limit and 10\% below the high-density limit.  The values of the line ratios corresponding to those density ranges have been obtained with the {\sc PyNeb} method {\tt Atom.getDensityRange}.}, the first one being efficient between 10$^{3.4}$ and 10$^{5.2}$ cm$^{-3}$, while the second one is sensitive between 10$^{4.0}$ and 10$^{5.8}$ cm$^{-3}$.
This implies that those two diagnostics are in principal not usable for low density regions. Nevertheless, the density determined by the 3 diagnostics being very similar, we will use n$_\mathrm{e}$(\forb{S}{ii}) for the low ionization region, and a mean of n$_\mathrm{e}$(\forb{Cl}{iii}) and n$_\mathrm{e}$(\forb{Ar}{iv}) for the high ionization region.

\begin{figure*}
\begin{center}
\includegraphics[width=0.90\linewidth]{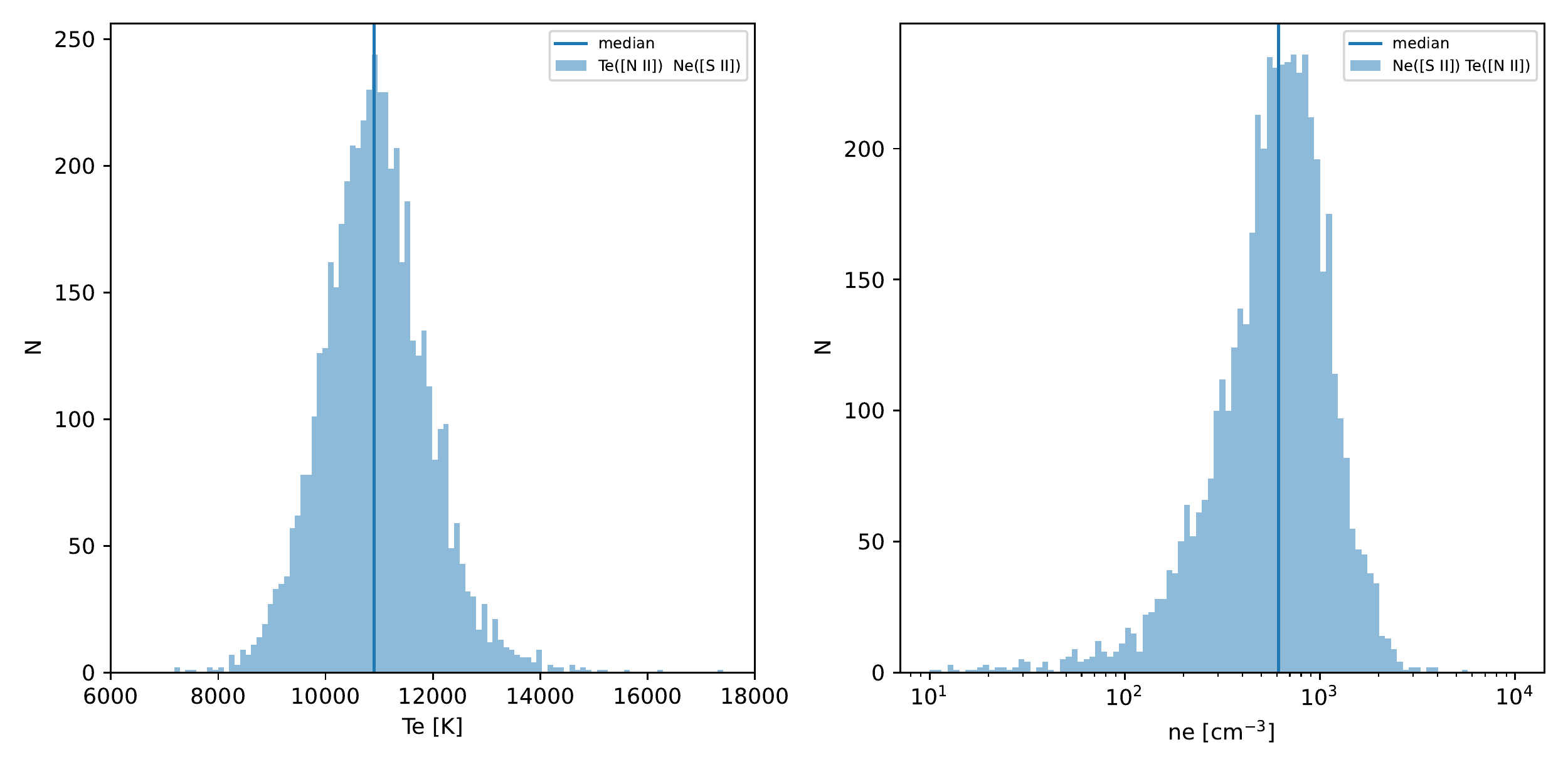}
\includegraphics[width=0.90\linewidth]{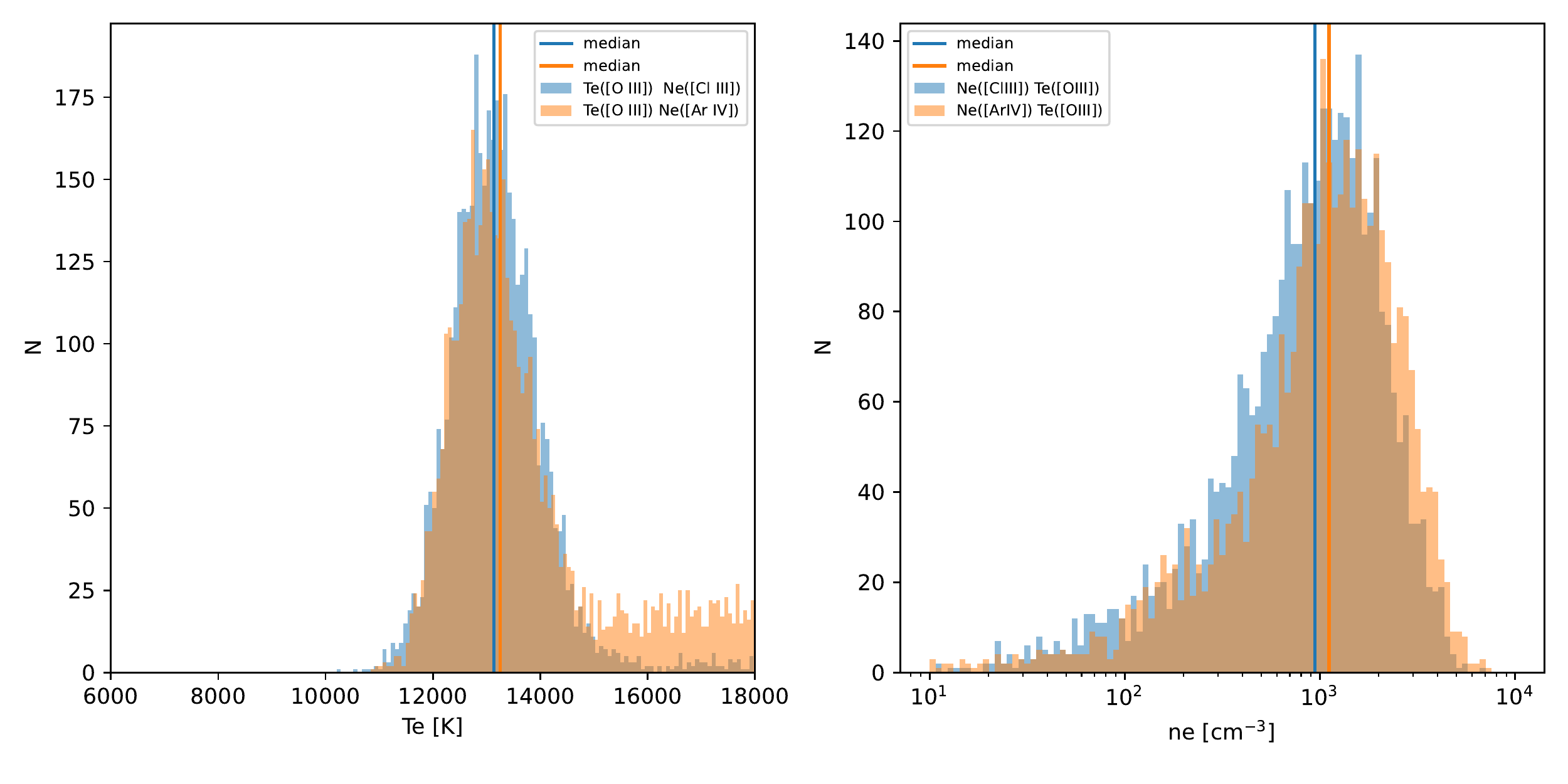}
\caption{Distribution of T$_\mathrm{e}$ (left panels) and n$_\mathrm{e}$ (right panels) using the Monte Carlo set of observations (see text). Upper panels: values obtained combining \forbr{N}{ii}{5755}{6584} for T$_\mathrm{e}$ and \forbr{S}{ii}{6716}{6731} for n$_\mathrm{e}$. Lower panels: values obtained combining \forbr{O}{iii}{4363}{5007} for T$_\mathrm{e}$ and \forbr{Cl}{iii}{5538}{5518} (blue) or \forbr{Ar}{iv}{4740}{4711} (orange) for n$_\mathrm{e}$.  Vertical lines show the median of each distribution.}
\label{TeNe}
\end{center}
\end{figure*}

For the MC distribution, we indicate in Table~\ref{diagnostics} the mean, median and standard deviation of the distribution. As aforementioned, the values obtained by applying the method to the original observed intensities is called "original" method. In Figure~\ref{TeNe} we show the MC distribution and its median value for the electron temperature and density. We observe that in the case of the electron density, the distribution is more symmetrical when using a logarithmic scale. 

This procedure therefore indicates that the electronic temperatures obtained with the mean and median are relatively equal to the value obtained directly with the observed line ratios. In all cases the standard deviation is small. Interestingly, we notice a high temperature tail in the values determined from the pair \forb{O}{iii} - \forb{Ar}{iv} (Fig.~\ref{TeNe}, bottom left): they come from MC realisations corresponding to low density. In the following, we will consider as the temperature for the high ionization zone the mean of T$_\mathrm{e}$ obtained from the \forb{O}{iii} - \forb{Cl}{iii} and \forb{O}{iii} - \forb{Ar}{iv} pairs. 
Once this mean value is obtained, the tail at high T$_\mathrm{e}$ is strongly reduced and does not significantly affect our main results.

The electronic densities derived with the MC method diverge slightly from the "original" method for the highly ionised species. Indeed, we can see in Figure~\ref{TeNe} and Table~\ref{diagnostics} that the distributions for \forb{Cl}{iii} and \forb{Ar}{iv} are biased toward higher values than the ones obtained directly from the observed line ratios. 
This is due to the fact that those two diagnostics are less efficient in the case of low density gas. This effect would be even more important in case of higher uncertainties on the line ratios. 
In all the cases the MC and "original" methods are in very good agreement with n$_\mathrm{e}$ $\sim $ 700~cm$^{-3}$ within the errors.

\begin{table*} 
\caption{T$_\mathrm{e}$, n$_\mathrm{e}$ and ionic abundances derived for the central area. For the Monte Carlo analysis we present the mean, median and standard deviation (std) for the different calculations.} 
\label{diagnostics}
\scalebox{1.0}{
\begin{tabular}{@{\extracolsep{4pt}}lcccc} \hline \hline 
\multicolumn{1}{|l|}{} &\multicolumn{1}{|r|}{{\bf Original}} & \multicolumn{3}{c}{{\bf Monte Carlo}}\\
\cline{2-2} \cline{3-5}
Diagnostic                      &         &Mean &  Median &  Std     \\
\hline 
\hline 
\input{table_tene.tex}
\hline 
\hline 
 Ionic abundances            &  Obs  & Mean   &    Median & Std           \\
\hline 
\hline 
\input{table_ionic.tex}
 \hline 
\hline
\end{tabular} }                                                                        
\end{table*}                                                                           
 
\subsubsection{Ionic abundances}

\begin{figure*}
\begin{center}
\includegraphics[width=.8\linewidth]{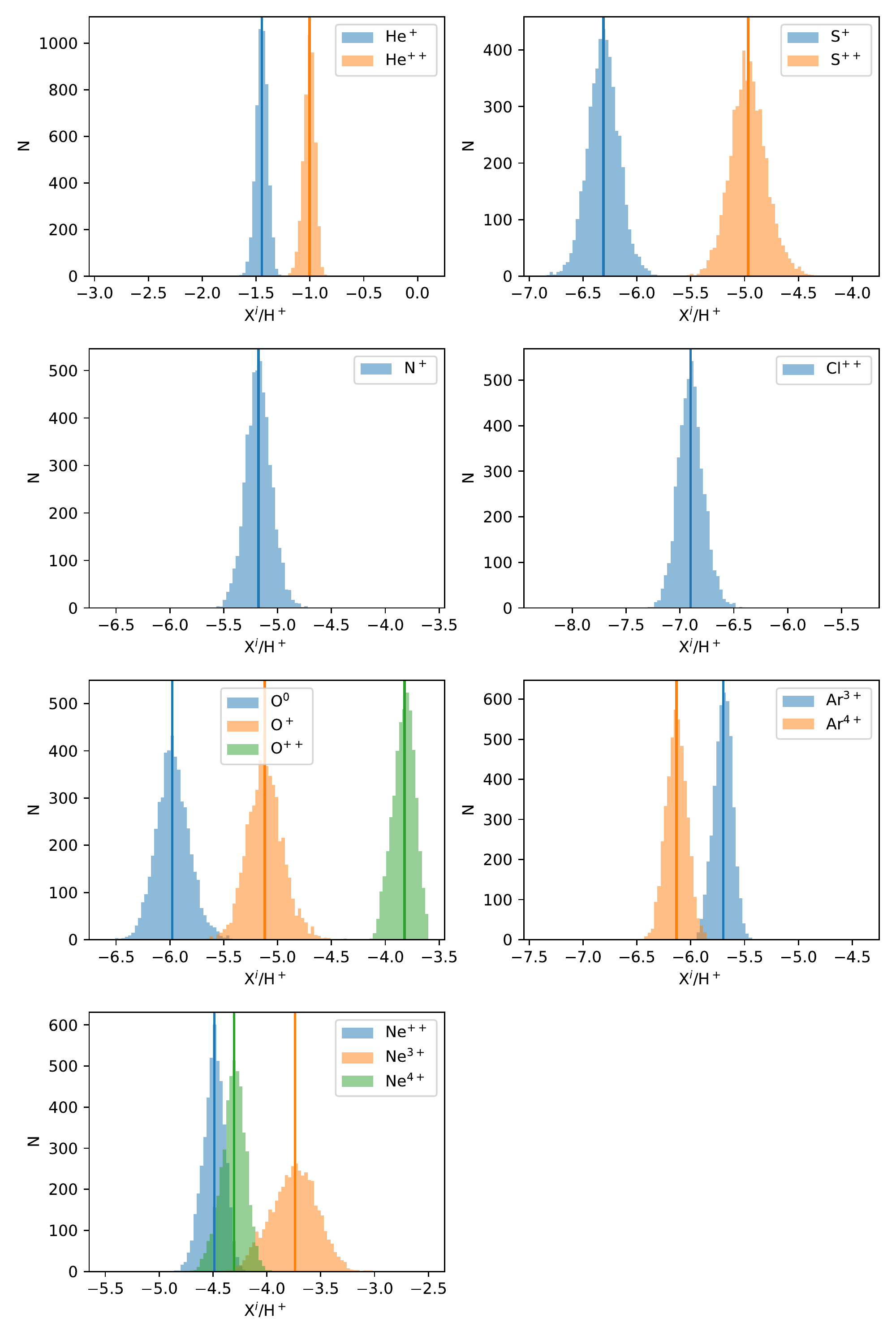}
\caption{Monte Carlo distribution of the ionic species. All the x-axis span over the same size. Vertical lines show the median of each distribution.}
\label{ionicabun1}
\end{center}
\end{figure*}

The ionic abundances are calculated with the adopted T$_\mathrm{e}$\,\forb{N}{ii} or T$_\mathrm{e}$\,\forb{O}{iii} based on the ionization potential, 30~eV being the threshold between low and high excitation regions (see Table~\ref{diagnostics}). Similarly to the electronic temperature and density, we proceeded to estimate the ionic abundances using the standard method as well as a MC procedure (with the corresponding temperatures and densities). The results are shown in the lower part of Table~\ref{diagnostics} and the MC distributions for each element are presented in Figure \ref{ionicabun1}.
These ionic abundances have been used to determine the total abundances as described in the next section.

\subsubsection{ICFs adapted to PC\,22 using Machine Learning}
\label{sec:ICFs}

To determine the elemental abundances from the ionic ones, the ionic abundances obtained from the observations are added and then a correction factor for the unseen ions is applied, the so-called Ionisation Correction Factor (ICF). The first ICFs were derived from similarities in the ionisation potentials of different elements \citep[e.g. N+/O+ = N/O, from][]{Peimbert1969}. ICFs can also be computed using photoionisation models (e.g. \citealt{KB1994}, hereafter KB94). More recently, \citet{DIMS2014} (hereafter DIMS14) computed a new set of ICFs from the large 3MdB database of models\footnote{ \url{https://sites.google.com/site/mexicanmillionmodels/}} \citep{Morisset2015}. 

In this work we will use the classical ICFs from KB94 as well as those of DIMS14. But we will also, {\it and for the first time}, use Machine Learning techniques to determine ICFs using the 3MdB grid of photoionisation models and to compute new ICFs specifically dedicated to our object (see a also our discussion in \S\ref{sec_disc_MLA}).
We choose to look for ad-hoc ICFs for our object because PC22 is at the limit where the DIMS14 ICFs are said by the authors not to be valid, the nebula being too highly ionised: since the ICFs based on O$^+$ are not usable here, we determine ICFs for chlorine and sulfur based on O$^{++}$. We also need ICFs to be applied to (Ar$^{3+}$ + Ar$^{4+}$) / H$^+$.

The main idea behind the determination of ICFs is to find photoionisation models that reproduce what is observed for a given object, in terms of ionisation stage, to predict the ionisation of other elements. Most of the time, one ionic fraction determined from the observation is used as a proxy to determine the ionisation stage of other elements, and {\it in fine} the ICFs. Sometimes 2 fractions are used, as in the ICFs by DIMS14, based on $v = $He$^{++} / ($He$^+ + $He$^{++})$ and $\omega=$O$^{++} / ($O$^+ + $O$^{++})$. But only two of their ICFs (namely N/O and Ne/O) are based on a combination of $v$ and $\omega$, the other ICFs refering to only one.
This is mainly due to the difficulty in finding analytic functions that describe the behaviour of one ionic fraction relative to the other one. This is basically a problem of multivariate regression, where Machine Learning based regressors excel. 

In our observations of PC\,22, we have 5 ionic abundances ratios that can be used to determine ICFs: He$^{++} / $He$^+$, O$^{++} / $O$^+$, Ne$^{4+} / $Ne$^{3+}$, Ne$^{3+} / $Ne$^{++}$, and Ar$^{4+} / $Ar$^{3+}$. The dominant ion Ne$^{3+}$ abundance is determined through the faint auroral line at 4726\AA, but we still prefer to rely on this ion to compute the ICF for Ne rather than the Ne$^{4+} / $Ne$^{++}$ ratio which implies two residual ions with large uncertainty.
We choose to compute the ICFs directly from the corresponding line ratios, instead of using ionic abundance ratios. We therefore need to have some constraints on the electron temperature that connect the line intensities to the ionic abundances: we choose to add the [OIII] 4363/5007 line ratio to the inputs of the Machine Learning (ML) process.

We use the python implementation of XGBoost \citep{Chen2016} to build the ML algorithm (MLA), using a learning rate of 0.1, a number of estimators of 500, and a maximum depth of 10 (we checked that changing these values to lower learning rates and higher number of estimators does not significantly changes the results, but increases the training time). Other hyper-parameters have their default values. The XGBoost library is called through the AI4neb facility (Morisset et al., in prep).

The input vector $X$ is build from a 6D vector of the logarithmic values of the following line ratios:
\begin{itemize}[topsep=0pt]
    \item \perml{He}{ii}{4686} / \perml{He}{i}{5876}
    \item \forbl{O}{iii}{5007} / \forbl{O}{ii}{3727} 
    \item \forbs{Ne}{v}{3426}{3346} / \forbl{Ne}{iv}{4726}
    \item \forbl{Ne}{iv}{4726} / \forbl{Ne}{iii}{3869}
    \item \forbl{Ar}{v}{6435} / \forbs{Ar}{iv}{4711}{4740}
    \item \forbr{O}{iii}{4363}{5007} 
\end{itemize}
We note that throughout the paper the notations \forbs{Ne}{v}{3426}{3346} and \forbs{Ar}{iv}{4711}{4740} indicate the sum of the emission lines.\\
The output vector $y$ is directly the set of the following ICFs (logarithmic values are used): 
\begin{itemize}[topsep=0pt]
    \item O / (O$^+$ + O$^{++}$)
    \item N/O $\times$ O$^+$ / N$^+$
    \item Ne / (Ne$^{++}$ + Ne$^{4+}$)
    \item Ne / (Ne$^{++}$ + Ne$^{3+}$ + Ne$^{4+}$)
    \item Ne / O $\times$ O$^{++}$ / Ne$^{++}$
    \item S / (S$^+$ + S$^{++}$)
    \item S / O $\times$ O$^{+}$ / (S$^+$ + S$^{++}$)
    \item S / O $\times$ O$^{++}$ / (S$^+$ + S$^{++}$)
    \item Cl / O $\times$ O$^{+}$ / Cl$^{++}$
    \item Cl / O $\times$ O$^{++}$ / Cl$^{++}$
    \item Ar / (Ar$^{3+}$ + Ar$^{4+}$)
\end{itemize}

XGBoost MLA is only able to predict one output. We then build 11 XGB MLAs, each one dedicated to the prediction of one of the ICFs from the output.

The MLAs are trained using a subset of the 3MdB database, corresponding to the PNe\_17 models (updated version of the database described in \citet{Morisset2015} and in the web site \url{https://sites.google.com/site/mexicanmillionmodels/}, using Cloudy v17.02).
We only consider models close to PC\,22 i.e. the models for which the predicted values differ from the observed values by less than 0.6 dex for each one of the 6 line ratios considered in the $X$ vector. These filters reduce the number of models from more than 720,000 (the whole 3MdB PNe database) to $\simeq$ 16,000. We then build 11 ad-hoc regressors locally around the position of PC\,22 in the $X$ space. The training is performed on 80\% of this set, the remaining 20\% being used for testing the performances of the MLAs. 

\begin{figure*}
\begin{center}
\includegraphics[width=.65\linewidth]{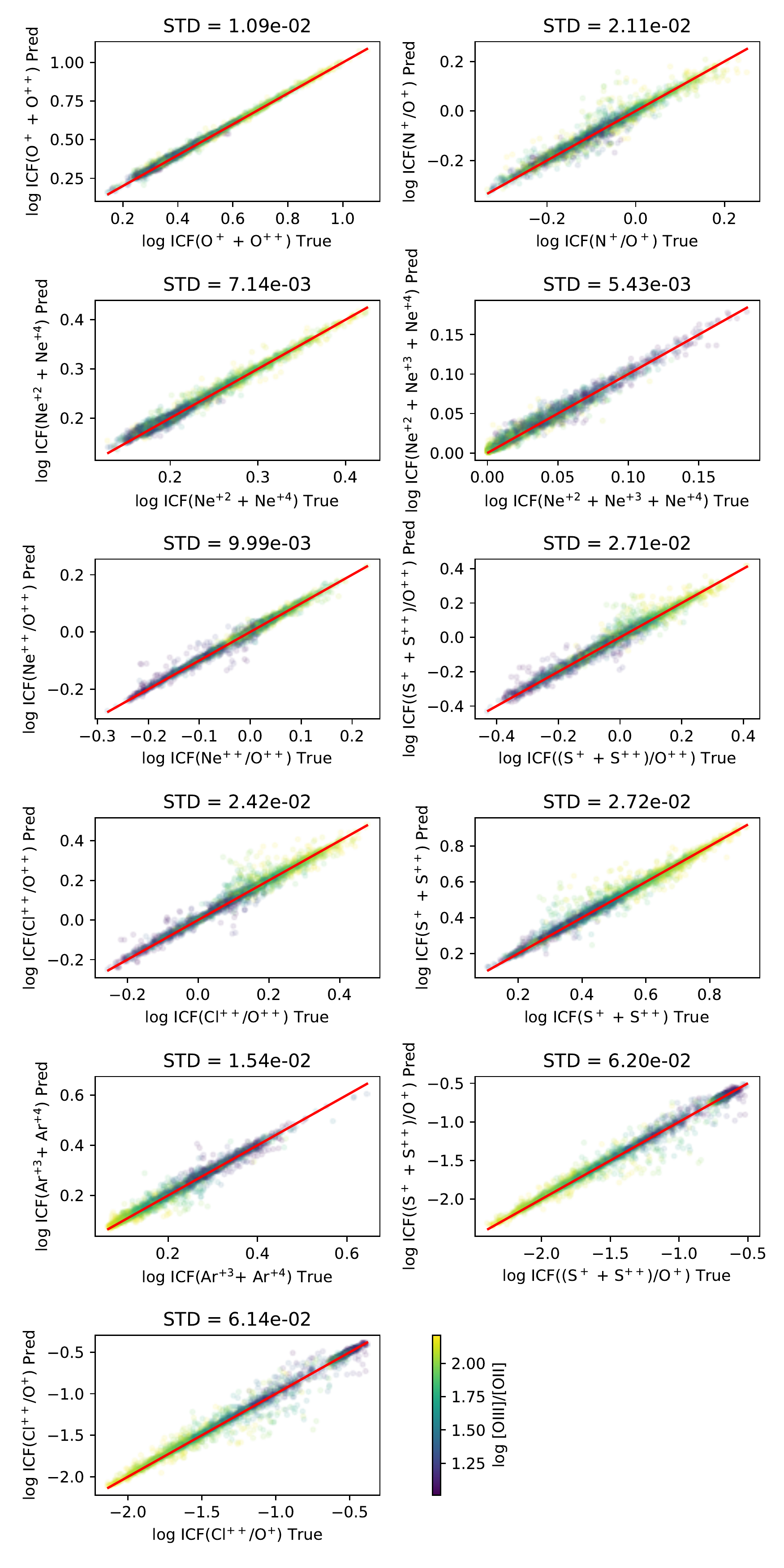}
\caption{\label{fig:icfs} Comparison between the predicted (on the y-axis) and the original values (on the x-axis) for the 11 ICFs used in this paper.}
\end{center}
\end{figure*}

The performances of the MLAs are shown in Figure~\ref{fig:icfs} where the predicted values of the 11 ICFs are compared to the real values, for the testing set. For each ICF, the value of the standard deviation of the distribution is given in dex above each plot. The color bar (the same is used for the 11 subplots) ranges over the values of log(\forb{O}{iii}/\forb{O}{ii}) selected for the training set, $\pm 0.6$ around the observed value of 1.6. For most of the cases, the regression is very satisfactory, with a few percents of deviation from the real value. The most problematic cases are the one normalized to O$^+$, which is a residual ion in these high excitation models. 

The analysis of the feature importance of the XGBoost trained MLAs shows that the most important line ratio is systematically \forb{O}{iii}/\forb{O}{ii}. The second most important is \perm{He}{ii}/\perm{He}{i} (in case of predicting ICF for O / (O$^+$ + O$^{++}$),  Ne / (Ne$^{++}$ + Ne$^{3+}$ + Ne$^{4+}$) and Ar / (Ar$^{3+}$ + Ar$^{4+}$) ) or \forb{Ne}{v}/\forb{Ne}{iv} for the other ICFs.

\begin{figure*}
\begin{center}
\includegraphics[width=.65\linewidth]{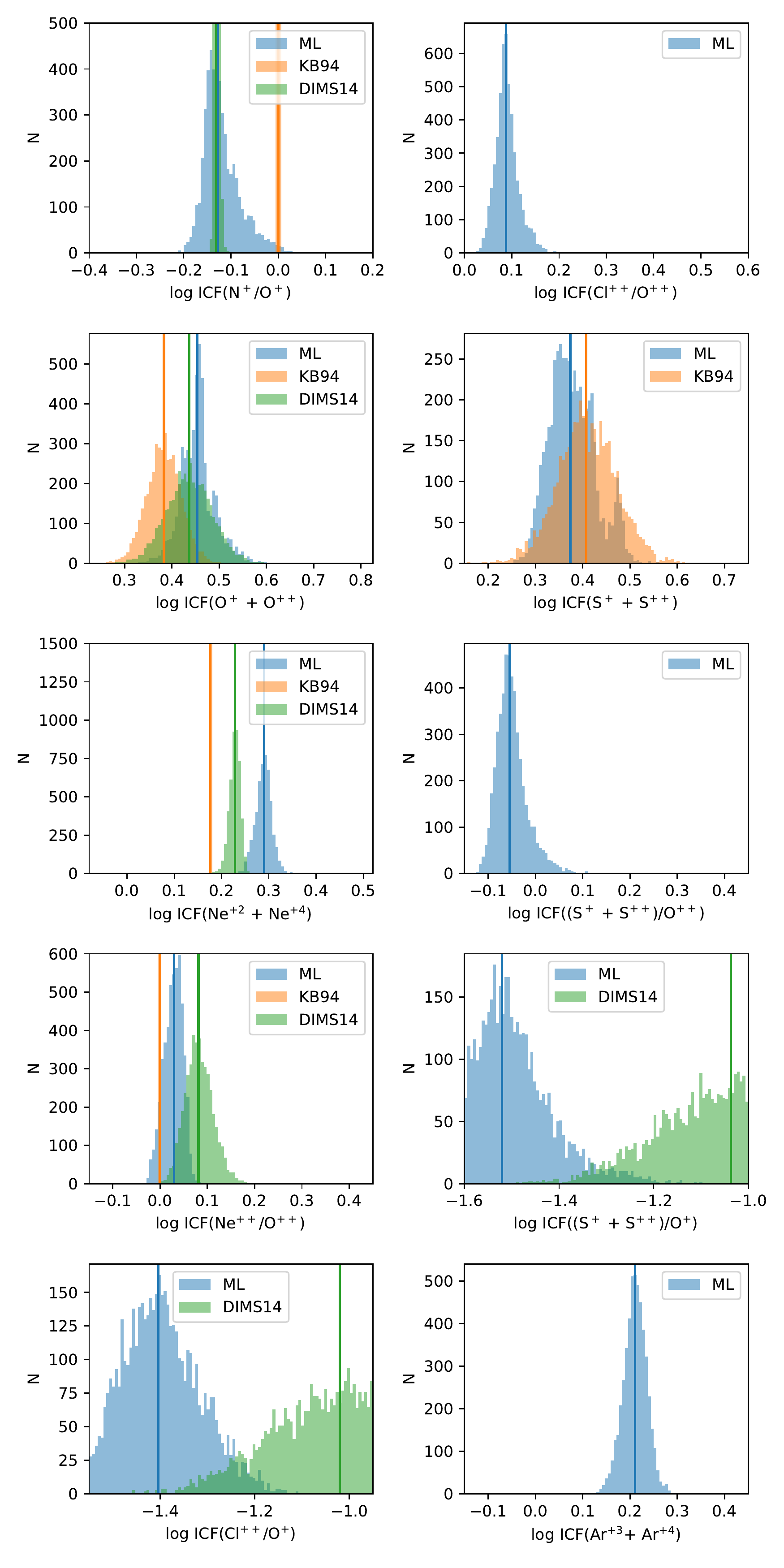}
\caption{\label{fig:icfs3} Determination of the ICFs to be used for the determination of the elemental abundances of PC\,22 using MLA, KB94 and DIMS14. All the x-axis span over the same size. Vertical lines show the median of each distribution.}
\end{center}
\end{figure*}

Once the MLAs are trained and tested, we can apply it to the determination of the ICFs for our observation and its 5,000 MC realisations. We also computed the ICFs obtained following KB94 and DIMS14. The results of the corresponding distributions are shown in Figure~\ref{fig:icfs3}. For some ICFs KB94 gives only one single value that does not depend on any line ratio, in this case no distribution is derived. {\it For (S$^{+}$ + S$^{++}$)/O$^{++}$, Cl$^{++}$/O$^{++}$, and Ar$^{3+}$+Ar$^{4+}$, we are producing pioneer ICFs, no comparison is possible with previous determinations}. 

The ICF applied to Ne$^{++}$ + Ne$^{3+}$ + Ne$^{4+}$ (not shown here) is very close to 1.0. The MLAs not being trained to be strictly positive, some values lower than 1.0 are predicted: they are set to 1.0 before using this ICF.

Once we have a good method to obtain ICFs dedicated to our object, we can proceed in the determination of the total abundances of He, N, O, Ne, Ar, Cl, and S.

\subsubsection{Total abundances}
\label{sec:elem_abund}

\begin{figure*}
\begin{center}
\includegraphics[width=.8\linewidth]{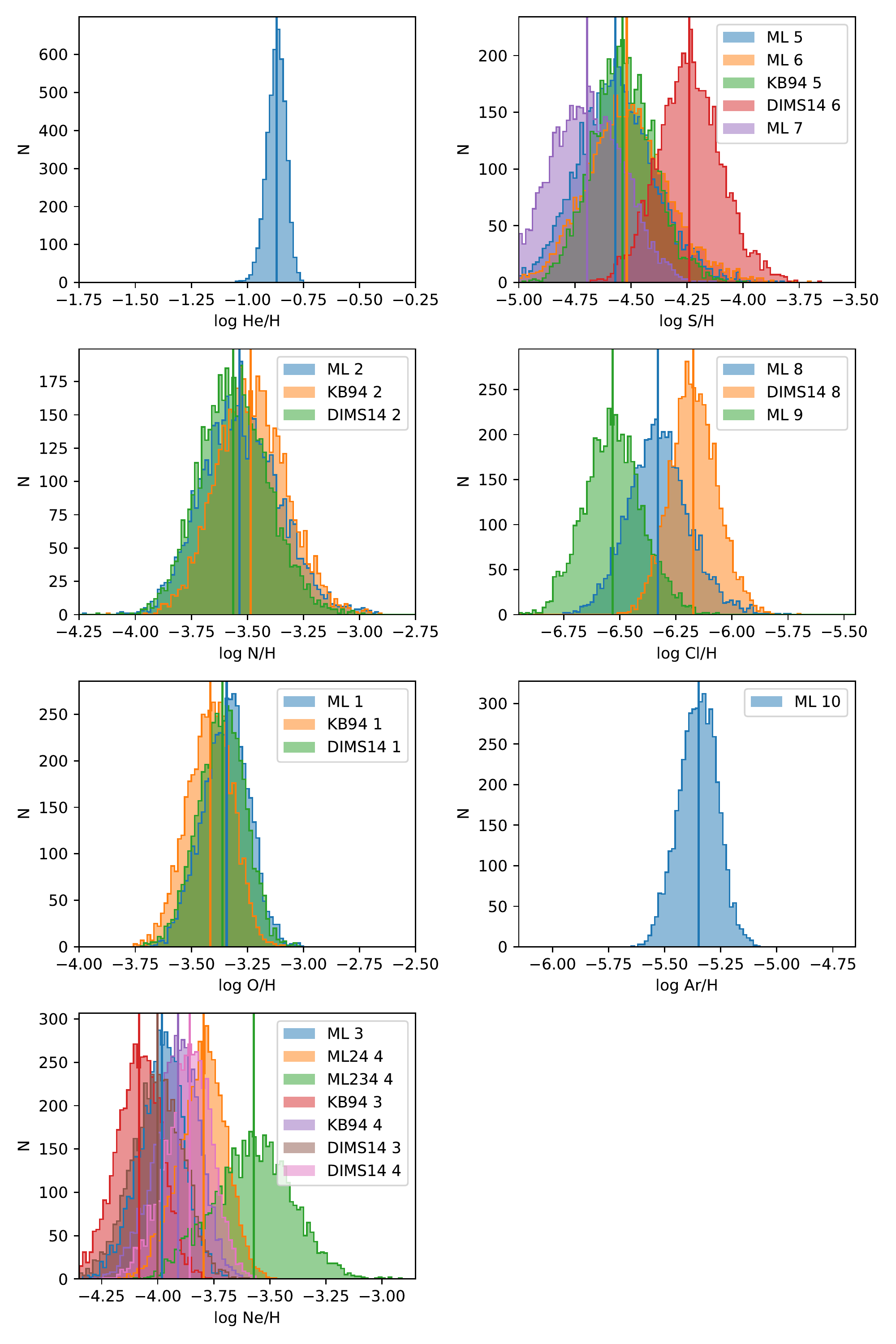}
\caption{\label{fig:abunds} Total abundances derived with the 3 methods for the ICFs (MLA, KB94 and DIMS14). See Tab.~\ref{tab:icf_abunds} for the labels of the ICFs. All the x-axis span over the same size. Vertical lines show the median of each distribution.}
\end{center}
\end{figure*}

We applied the ICFs determined in the previous section to the ionic abundances to obtain the element abundances. The corresponding distributions are shown in Figure~\ref{fig:abunds}. As in the case of the ICFs presented in the previous section, for each method used to obtain abundances, the value corresponding to the median of the distributions is shown with a vertical line. These values, as well as the "original" value, the mean value and the standard deviations of the distributions, are reported in the lower part of Table~\ref{tab:icf_abunds}.

The difference in using MLA, KB94 or DIMS14 for the determination of the nitrogen or oxygen abundance is small (less than 0.1 dex). 
In the case of the sulphur abundance, our result is closer to that obtained from the ICFs of KB94 (0.06 dex) than that from the ICFs of DIMS14 (0.25 to 0.40 dex from MLA, depending on the ICF used.)
As explained earlier, {\it only our estimation of the argon abundance is present} which is a unique and interesting fact. 

The chlorine abundance derived with the MLA can only be compared to that derived with DIMS14's ICFs, and the difference is not negligible with $\approx$0.13 to $\approx$0.28 dex. 

Finally, an interesting feature is the determination of the neon abundance where there is a difference whether Ne$^{3+}$ is included in the calculation of the ICFs or not. KB94 and DIMS14 do not consider this line, which was likely too faint (or absent) to be taken into account. In our case, the use of Ne$^{3+}$ changes the constraint on the neon (as the ICF is now equal to 1) and this would imply that our estimation of Ne/H is more correct as all the neon ions are taken into account. However, while the line is detected in PC\,22, it is very faint ($\sim$2\% of H$\beta)$ and very sensitive to the electron temperature. Changing the value of T$_e$ from 13,000 to 15,000~K increases its emissivity by a factor of 2.6, reducing the determined value of Ne/H by almost the same factor. 
This could therefore cast some doubts on the precision of the derived abundance. Also, we have to take into account the error on the measurement of the neon abundances using the ML with the Ne$^{3+}$ included. 

\begin{table*} 
\caption{ICF and total abundances derived for the central area.} 
\label{tab:icf_abunds}
\scalebox{1.0}{
\begin{tabular}{@{\extracolsep{4pt}}lccccc} \hline \hline 
\multicolumn{1}{|l|}{} &\multicolumn{1}{|r|}{{\bf Original }} & \multicolumn{3}{c}{{\bf Monte Carlo}}\\
\cline{2-2} \cline{3-5}
ICF                      &          &Mean &  Median &  Std    & Ref \\
\hline 
\hline 
\input{table_icf.tex}
\hline 
\hline 
Total abundances            & Original & Mean   &    Median & Std  & Ref ICF          \\
\hline 
\hline 
\input{table_elem.tex}
 \hline 
\hline
\end{tabular} }                                                                        
\end{table*}

\section{Stellar analysis}\label{CS_analysis}

\subsection{Spectral analysis output}
\begin{table}
\begin{center}
\caption{Parameters for the emission lines considered for the multi-Gaussian fitting of the WR features of the CS.}
\label{tab:elines}
\setlength{\tabcolsep}{0.6\tabcolsep}
\begin{tabular}{ccccccc}
\hline
{ID} &    & Line         ({$\lambda_0$})& F    & FWHM &  EW  & $F/F_\mathrm{RB}$ \\
(1)  & (2)& (3,4)                    & (5)  &(6)   & (7)  & (8) \\
\hline
VB   &    &                           &      &      &      &      \\
1    & WR & O\,{\sc vi}   (3820)        &231.4 & 65.7 &105.7 &7561.3 \\
BB   &    &                           &      &      &      &       \\
2    & WR & He\,{\sc ii}  (4686)        & 24.2 & 27.1 & 19.5 & 789.7 \\
3    & WR & C\,{\sc iv}   (4658 )       & 25.1 & 27.1 & 19.7 & 819.9 \\
4    & Neb& He\,{\sc ii}  (4686 )       & 90.7 &  4.3 & 73.3 & --   \\
RB   &    &                           &      &      &      &      \\
5    & WR & C\,{\sc iv}   (5801)        &  1.5 &  4.2 &  2.1 & --   \\
6    & WR & C\,{\sc iv}   (5812)        &  1.5 &  9.2 &  2.1 & --   \\
other &   &                           &      &      &      &      \\
7    & WR & O\,{\sc vi}  (5290)        &  4.6 & 13.3 &  5.4 & 150.0   \\
8  & WR & Ne\,{\sc viii}  (6068)        &  1.5 &  6.2 &  2.7 & 50.3   \\

\hline
\end{tabular}\\
(1) Identification number of the Gaussian components in the WR features;
(2) Nature of the contributing emission line: WR (broad) or nebular (narrow);
(3) Line;
(4) Rest wavelength in \AA;
(5) Flux in units of $10^{-16}$~erg~cm$^{-2}$~s$^{-1}$;
(6) Full Width at Half Maximum (FWHM) [\AA];
(7) Equivalent Width (EW) [\AA];
(8) Line fluxes normalized to the RB (\perms{C}{iv}{5801}{5812}). The ratio has been computed adopting a $F$(RB)=100.
\end{center}
\end{table}

\begin{figure*}
\begin{center}
\includegraphics[angle=0,width=1.0\linewidth]{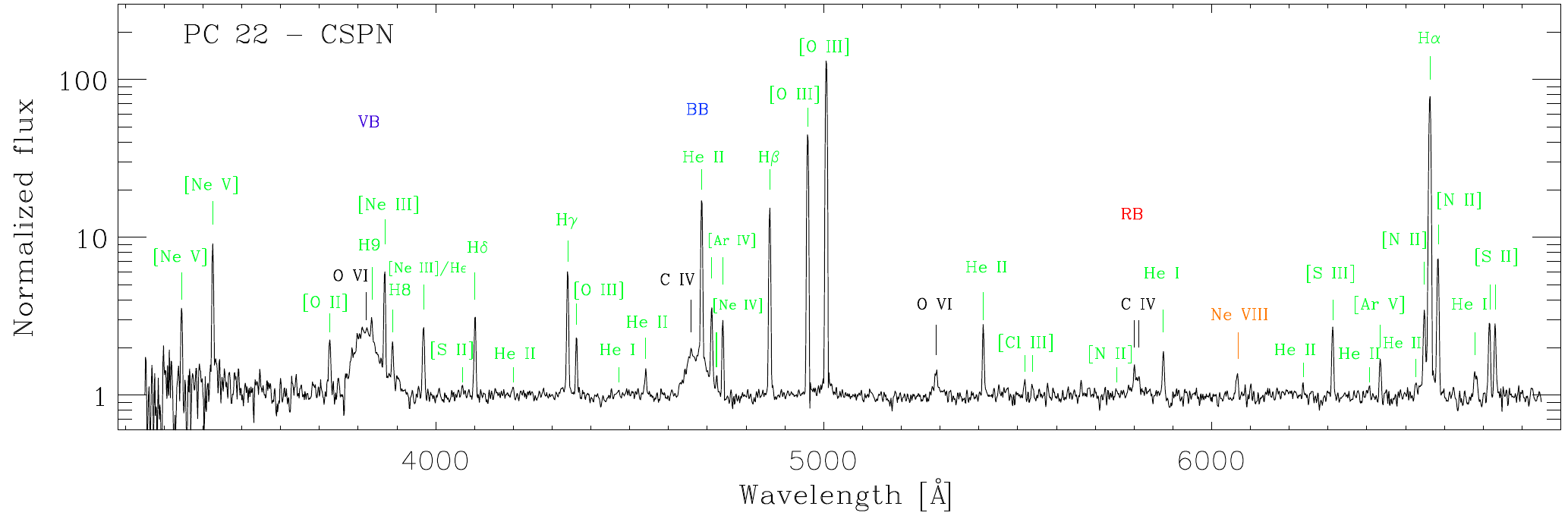}
\caption{NOT ALFOSC spectrum of the CSPN of PC\,22.
The violet bump (VB) at 3820~\AA, the blue bump (BB) at 4686~\AA\, and the red bump (RB) at 5806~\AA, the most common optical WR features, as well as the narrow lines from the nebular environment are indicated with black and green labels, respectively. The neon line of high ionization potential is indicated with orange. The spectrum is shown normalized to the best fit continuum spectrum.}
\label{fig:NGC2371_spec_star}
\end{center}
\end{figure*}

\begin{figure*}
\begin{center}
\includegraphics[angle=0,width=0.325\linewidth]{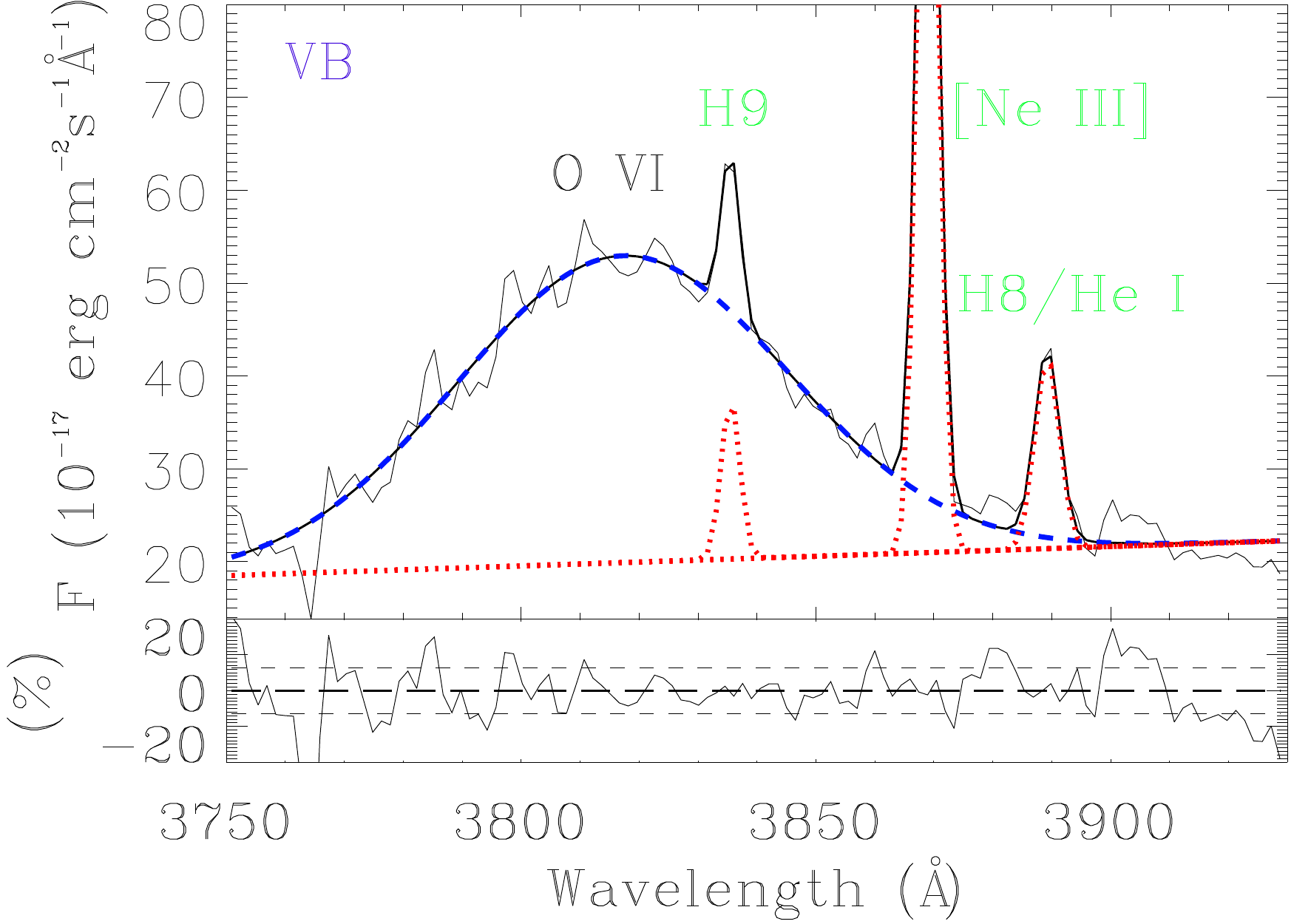}~
\includegraphics[angle=0,width=0.325\linewidth]{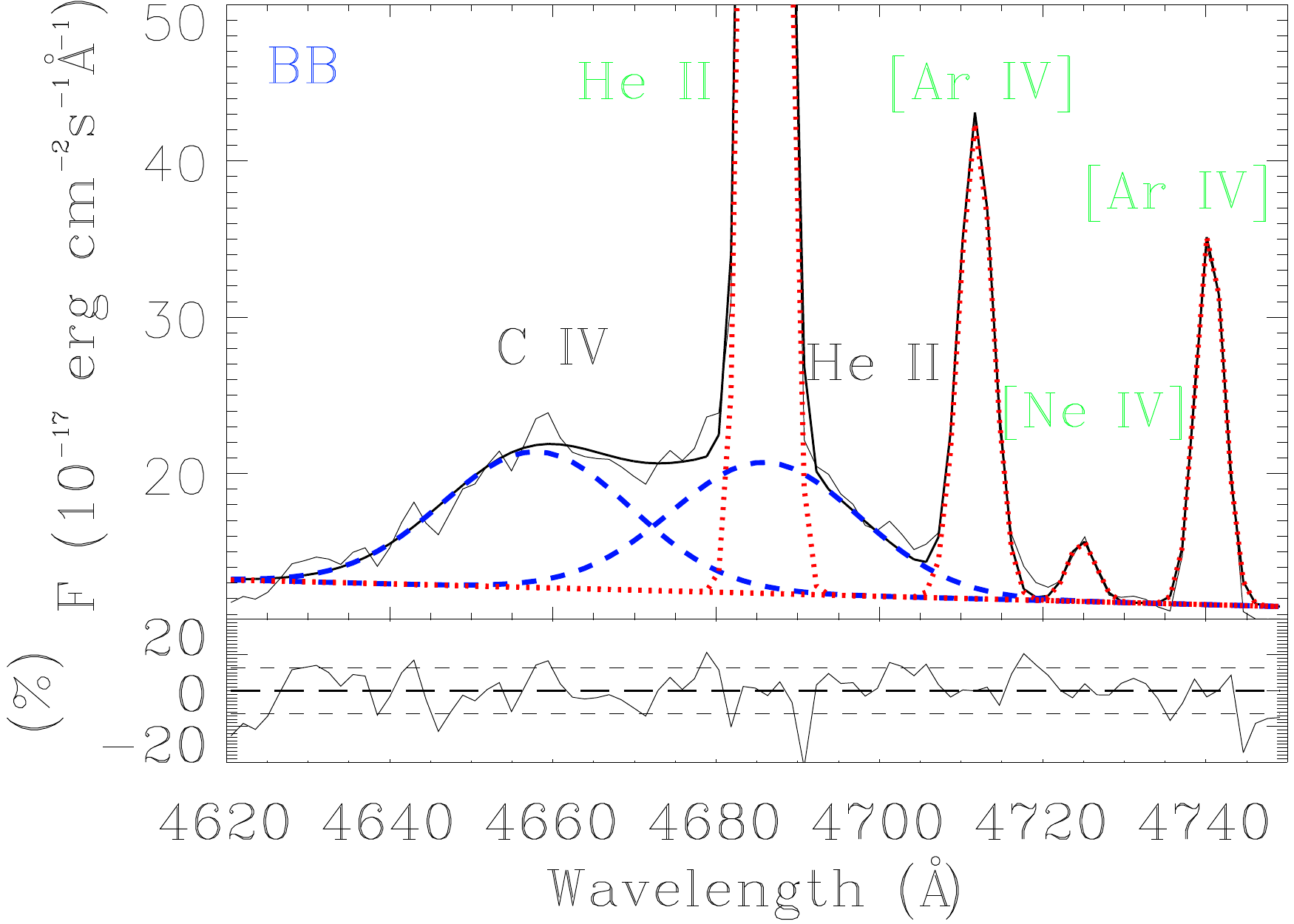}~
\includegraphics[angle=0,width=0.325\linewidth]{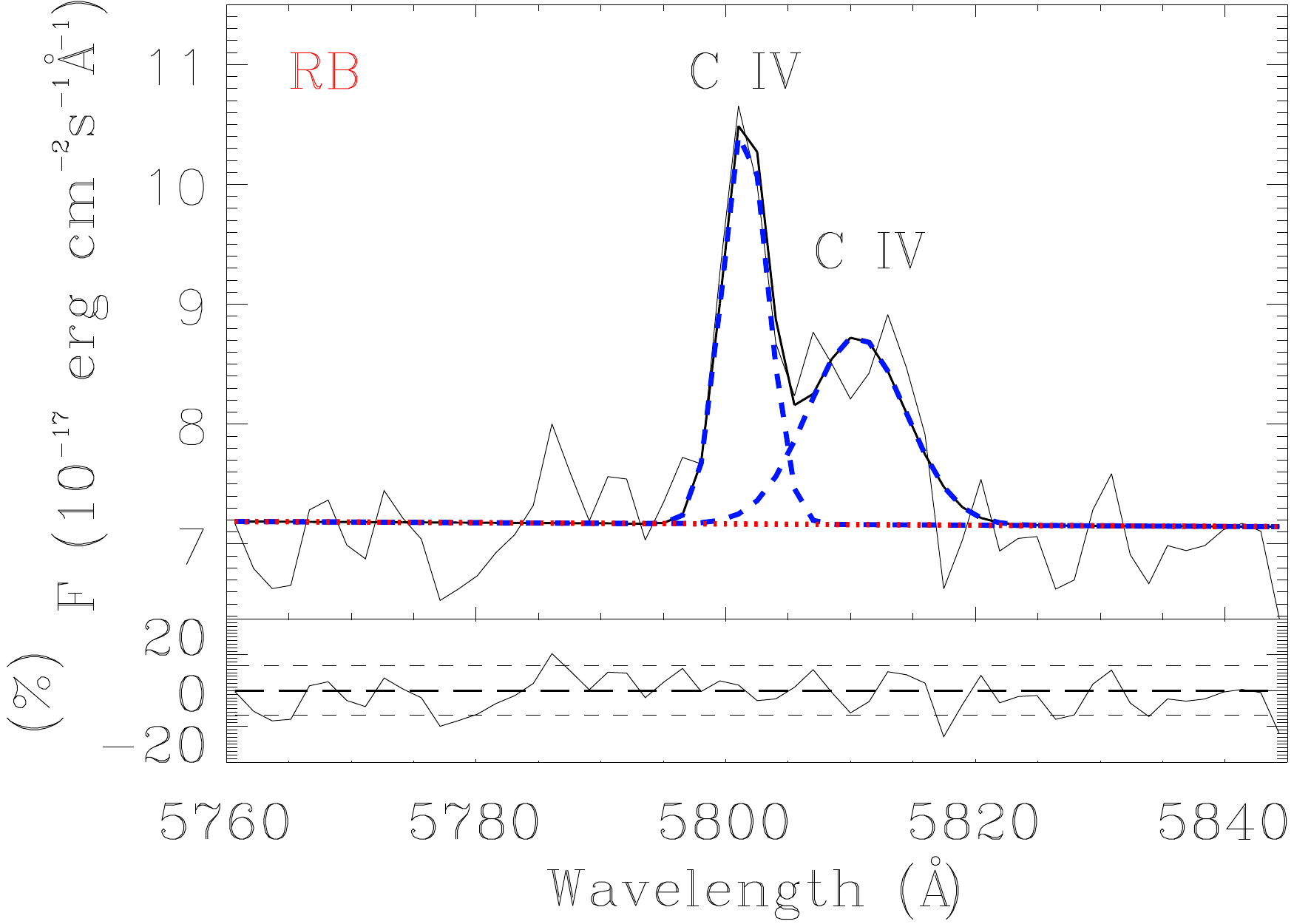}\\
\includegraphics[angle=0,width=0.325\linewidth]{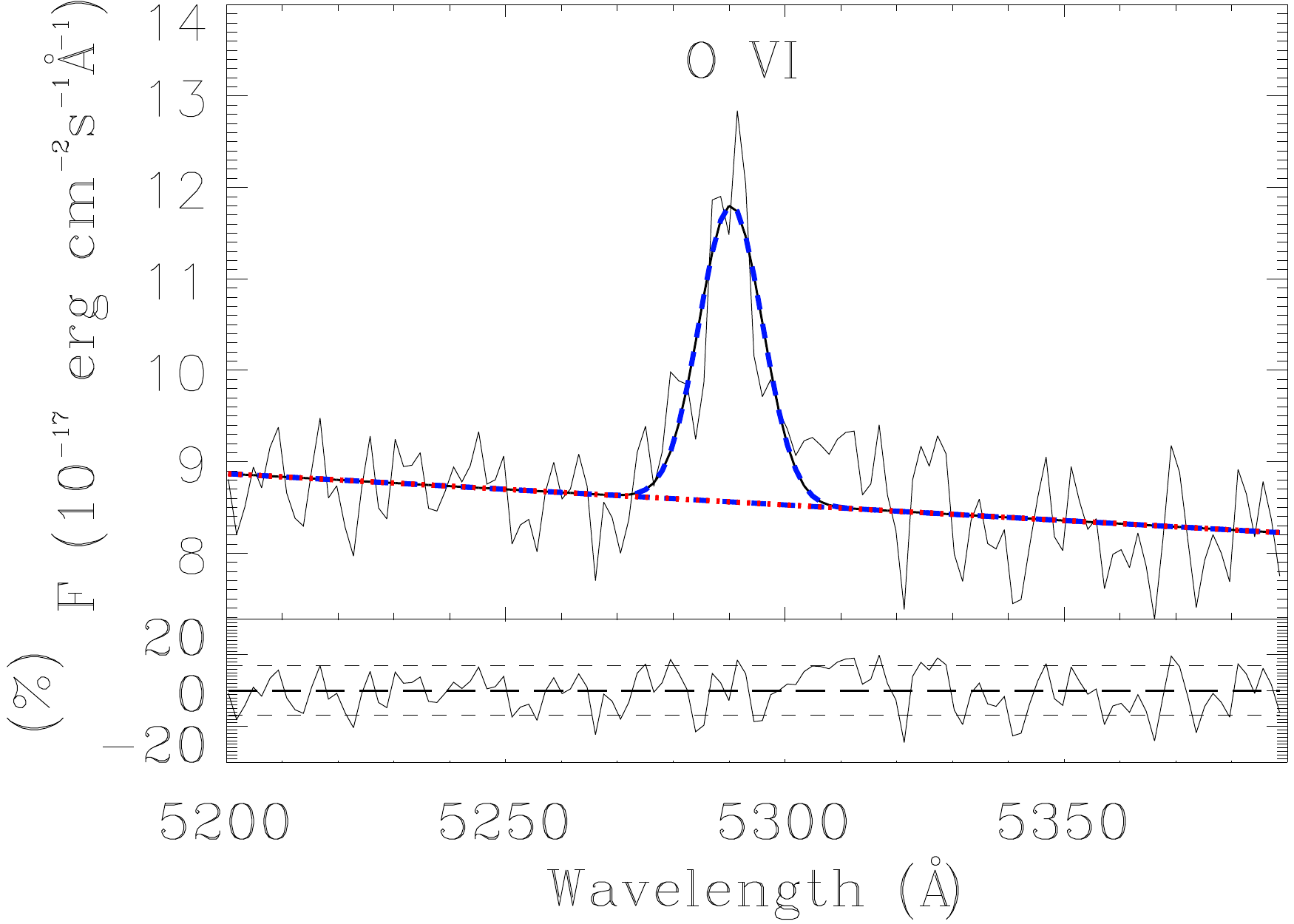}~
\includegraphics[angle=0,width=0.325\linewidth]{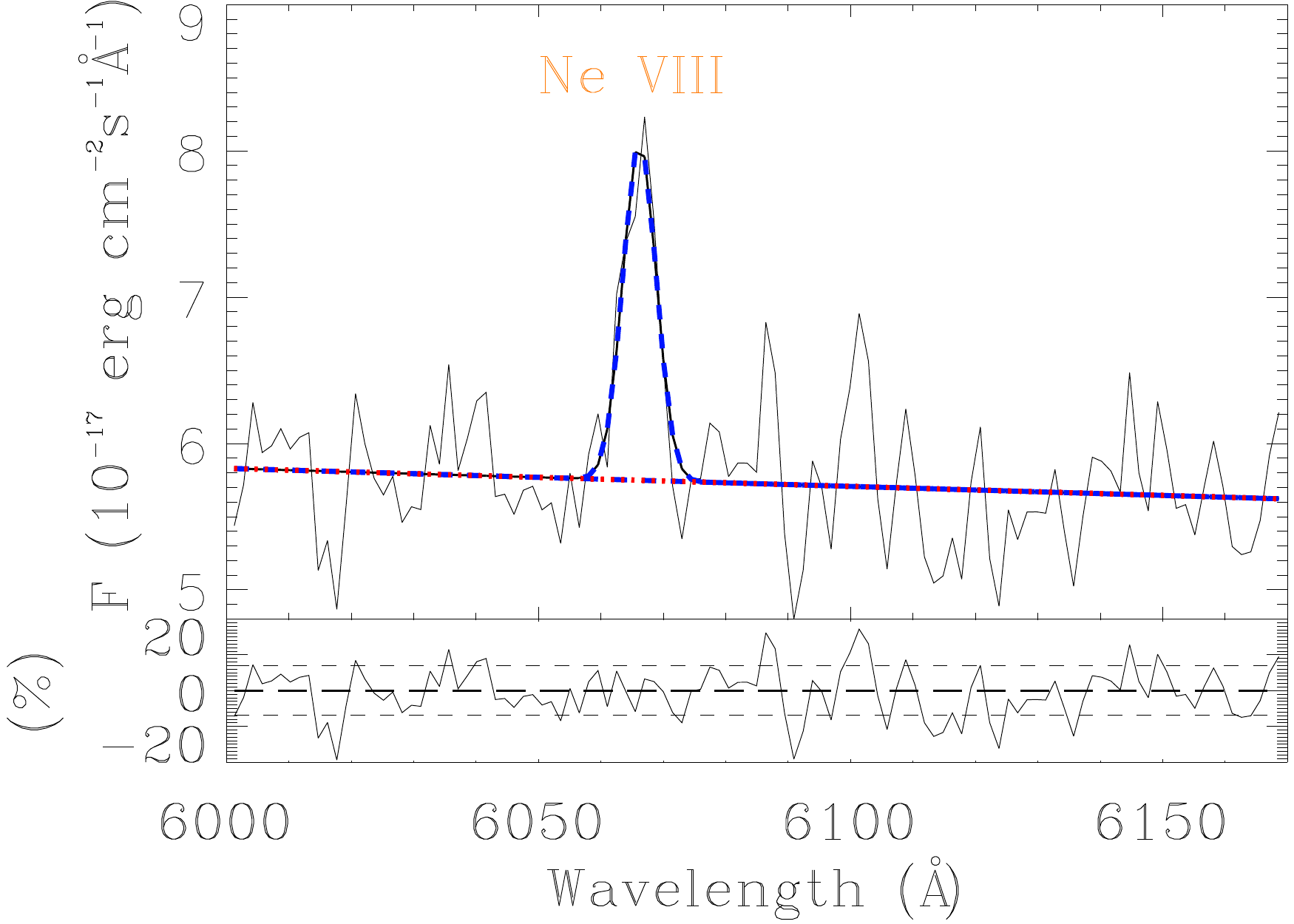}\\
\caption{Multi-component Gaussian fits to the WR features in the spectrum of the CSPN of PC\,22: the O-bump ({\it left}), the blue bump ({\it center}) and the red bump ({\it right}). The fitted components are indicated: the WR broad lines in {\it dashed blue}; the nebular lines in {\it dotted red}; the sum is shown in black. The fitted continuum is shown by the {\it dashed straight line}. The fitted WR features and nebular components are indicated with black and green labels, respectively. Residuals in per cent (\%) are shown at the bottom of each panel. The results are listed in Table~\ref{tab:elines}.}
\label{fig:gauss}
\end{center}
\end{figure*}


The spectrum of the central star of PC\,22 (Fig.~\ref{fig:NGC2371_spec_star}) exhibits several WR features,
including the so-called blue and red bumps (BB and RB) at $\sim$4686~\AA\ and
$\sim$5806~\AA, respectively, as well as the oxygen or violet bump (VB) at $\sim$3820~\AA.
Other emission lines related to WR features are also present in the
optical spectrum: \perml{O}{vi}{5290} and \perml{Ne}{viii}{6068}.

We note that the spectrum was corrected for extinction by
using the $c$(H$\beta$) value estimated from the Balmer decrement method.
We assume an intrinsic Balmer decrement ratio corresponding to a case B 
photoionised nebula of $T_{\rm e}$ = 10,000~K and $n_{\rm e}$ = 100~cm$^{-3}$ \citep[see][]{Osterbrock2006} and the reddening curve of \cite{Cardelli1989}.

The VB and BB are composed of several blended emission lines and
need a careful fitting in order to separate their broad and narrow
contributions.
This was done by applying the analysis described in detail by \cite{GomezGonzalez2020}.
The method consists on fitting the broad WR features with multi-Gaussian
components using a tailor-made code that uses the {\sc idl} routine
{\sc lmfit}\footnote{The {\sc lmfit} function (lmfit.pro) performs a non-linear
least squares fit to a function with an arbitrary number of parameters.
It uses the Levenberg-Marquardt algorithm, incorporated in the routine
{\it mrqmin} from \citet{Press1992}.}.
As a result, the fluxes, central wavelengths, FWHMs and equivalent widths (EW)
from the WR spectral features, as well as contributing nebular lines, have been estimated.
All these parameters are listed in Table \ref{tab:elines}.
We note that the FWHM of lines listed as WR features is larger than that of the nebular lines.

Figure \ref{fig:gauss} shows the fits for the different WR features of the central star of PC\,22.
The VB feature is made of a broad \perml{O}{vi}{3820} (FWHM$\sim$66~\AA) of stellar origin, 
and several narrow lines of nebular origin, such as H and Ne.
In the BB feature the broad \perml{He}{ii}{4686} WR line is blended with the \perml{C}{iv}{4658}.
In addition, the narrow emission lines \perm{He}{ii} and \forb{Ar}{iv},
present for highly excited nebulae, also contribute to the BB.
There is no contribution of nebular lines to the RB, made of the \perms{C}{iv}{5801}{5811} doublet.

The classification system concerning low mass [WR] stars is based on
the relative strength of the carbon (\perm{C}{ii}, \perm{C}{iii} and \perm{C}{iv}) and oxygen (\perm{O}{v}, \perm{O}{vi}, \perm{O}{vii} and \perm{O}{viii}) lines \citep{Acker2003}. We note that the last two features were reclassified by \citet{Werner2007} as \perm{Ne}{vii} and \perm{Ne}{viii} lines, respectively, from stellar origin.
First, an appropriate criterion to separate WO from WC appears to be
the absence of \perm{C}{iii}, as proposed by \citet{Kingsburgh1995}.
A WCL-type is excluded since the spectrum of the central star of PC\,22 does not show any \perml{C}{ii}{4267}
and/or \perml{C}{iii}{5696} (see Figure 1 and Table 3 in \citealt{Crowther1998}).
Without this line, the WC criteria fail.
%
The presence of the \perms{O}{vi}{3811}{3834} doublet is essential
for the classification as a [WO]-subtype.
The central star presents \perml{O}{vi}{3820} and \perml{O}{vi}{5290} additional to the
\perms{C}{iv}{5001}{5012} doublet. This is consistent with a [WO1]-subtype. 
Also, quantitatively, for a [WR] to be classified as [WO1],\, log(\perml{O}{vi}{3820}/\perml{C}{iv}{5806}) > 0.2, which applies for the spectrum of the central star as this translates into 

\begin{enumerate}

    \item $log(EW_\text{\perml{O}{vi}{3820}}/EW_\text{\perml{C}{iv}{5806}}) = 1.7$, ([WO1 $\geq0.2$; WO2 $\geq0.2$; WO3 = [$-1$ to 0.2]), which is closer to [WO1].

\end{enumerate}

Based on \citet[][]{Acker2003}, we performed a quantitative classification of the central star with the following criteria:

\begin{enumerate}

    \item $F_\text{\perml{O}{vi}{3820}}/F_\text{RB} = 7561.3$, ([WO1 $>1400$; WO2 = 1000$\pm$200; WO3 = 250$\pm$40), consistent with [WO1];
    \item $F_\text{\perml{C}{iv}{4658}}/F_\text{RB} = 819.9$, ([WO1 = $300\pm100$; WO2 = $270\pm60$; WO3 = $23\pm2$),  consistent with [WO1];
    \item $F_\text{\perml{C}{IV}{4686}}/F_\text{RB} = 789.7$, ([WO1 = $500\pm200$; WO2 = $300\pm30$; WO3 = $130\pm30$), consistent with [WO1];
    \item $F_\text{\perml{Ne}{vii}{5290}}/F_\text{RB} = 150.0$, ([WO1 $>80$; WO2 = 48$\pm$2; WO3 = 20$\pm$5), consistent with [WO1];
    \item $F_\text{\perml{Ne}{viii}{6068}}/F_\text{RB} = 50.3$, ([WO1 = $20\pm8$; WO2 = 6$\pm$1; WO3 = $2\pm1$), consistent with [WO1].

\end{enumerate}

We note that the FWHMs of the \perms{C}{iv}{5801}{5811} doublet are 4.2 and 9.2 \AA, respectively, which are below expected ([WO1 = 33$\pm$5; WO2 = 32$\pm$3; WO3 = 37$\pm$6). WCL are expected to have such small FWHMs, however, this classification is completely ruled out for our central star as it does not present any \perml{C}{ii}{4267}
and/or \perml{C}{iii}{5696} as stated above. Furthermore, only the early [WO]-types are expected to display strong \perml{O}{vi}{3820} broad emission lines.

Finally, the sole presence of \perml{Ne}{viii}{6068}, once thought to be \perm{O}{viii}, is enough to assign a temperature > 150,0000~K according to \citet{Werner2007}. We highlight that this line has only been observed in early-type [WO] central stars.
 
\begin{figure*}
\begin{center}
\includegraphics[angle=0,width=0.9\linewidth]{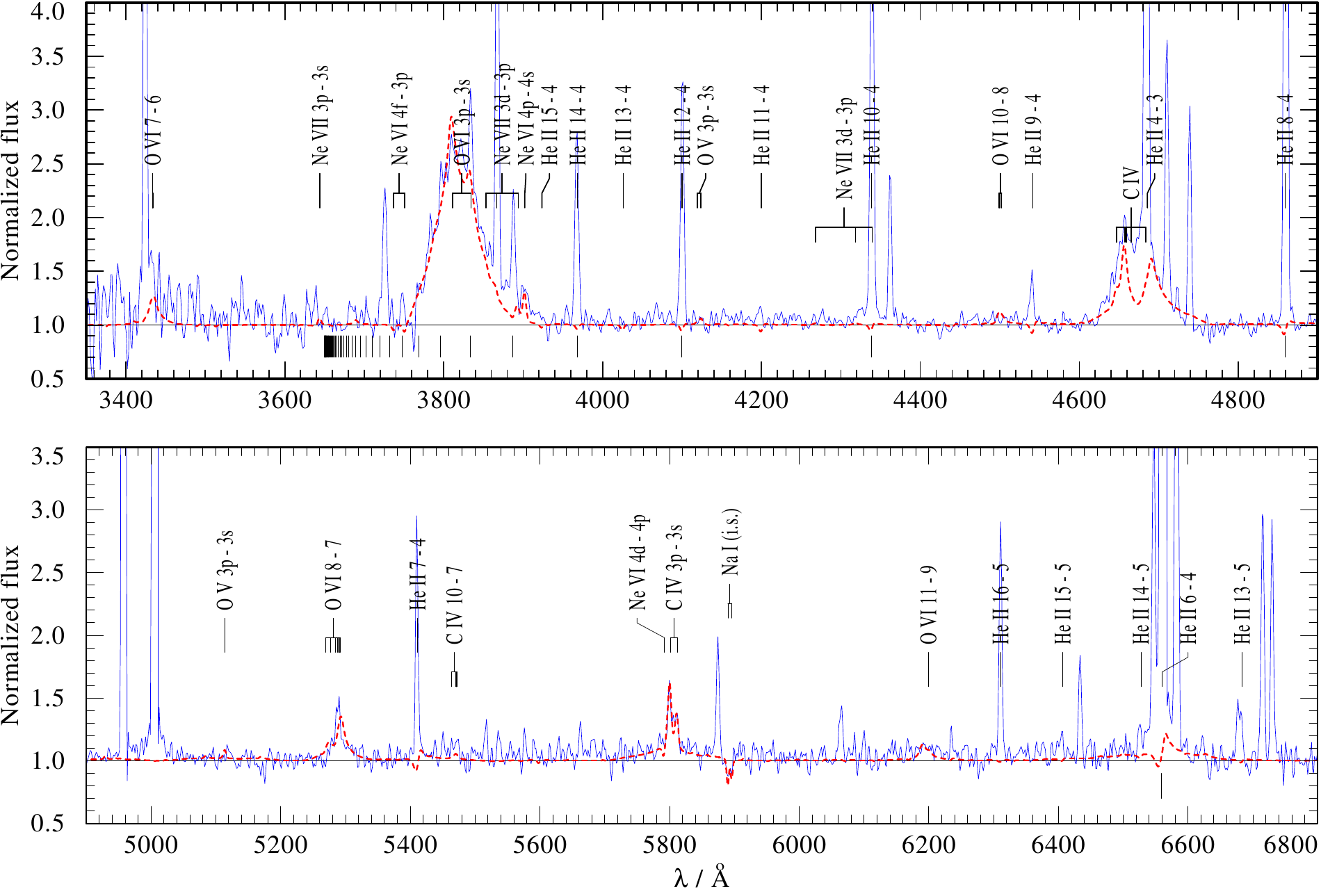}~
\caption{Detail of the optical spectrum: Best fitting PoWR model (red dashed line) vs. observation (blue solid line). The observation is normalized by the
model continuum. Relevant stellar lines are labeled. The
small vertical bars below the spectrum indicate the nebular Balmer lines.
For a better comparison the model spectrum is convolved with a Gaussian
of 3.8 \AA{} FWHM corresponding to the spectral resolution of the
observation, inferred from the interstellar Na I D doublet.
}
\label{fig:optical-model}
\end{center}
\end{figure*}

\begin{figure*}
\begin{center}
\includegraphics[angle=0,width=0.9\linewidth]{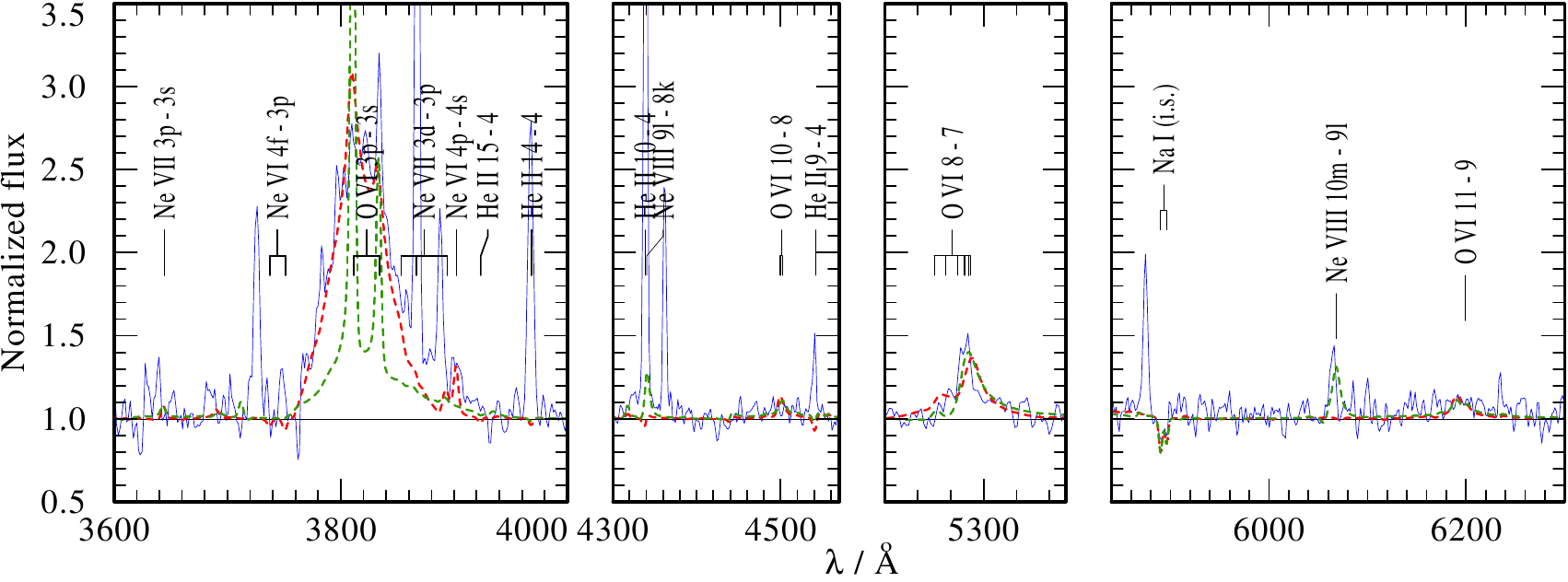}~
\caption{PoWR model with an increased stellar temperature of 178 kK
(dashed green) established in order to try to fit the \perml{Ne}{viii}{6068} line (see text). Note the impact of the higher temperature on the \perml{O}{vi}\, 3p-32
$\lambda$ 3811.4,3834.3 doublet while the the other \perml{O}{vi}\, lines are hardly changed.}
\label{fig:optical-ovi-neviii}
\end{center}
\end{figure*}

\subsection{Using 3MdB to determine T$_\mathrm{eff}$}

We also made use of 3MdB to estimate the effective temperature of PC 22. The main caveat is the use of a database in which the models are built with a black body as a model atmosphere to describe the central star, while we already established the [WR] nature of the nucleus of PC 22. Hence, owing to the fact that there is no available UV data for the source that would have allowed us to better constrain the nucleus characteristics, we will concentrate on {\it reproducing the ionization state of the gas} seen in PC 22.

We looked for the models in 3MdB that reproduce simultaneously \perml{He}{ii}{4686} / H$\beta$ and the consecutive ions line ratios: \forbl{O}{iii}{5007} / \forbl{O}{ii}{3727}, \forbl{Ar}{v}{6435} / \forbs{Ar}{iv}{4740}{4711}, \forbl{Ne}{iv}{4726} / \forbl{Ne}{iii}{3868}, and \forbs{Ne}{v}{3426}{3346} / \forbl{Ne}{iv}{4726}. The number of models coinciding with the ionization state of the PN depends on the tolerance used when defining the mask. Hence, we identified 4 models fitting the observations of PC 22 within a tolerance of 0.15 dex and 47 within a tolerance of 0.20 dex. In the former case the corresponding T$_{eff}$ range from 230,000 to 250,000~K, and in the latter case T$_{eff}$ range from 140,000 to 250,000 K. These values, although quite high, agree with the prediction by \citet{Werner2007} mentioned above.

\subsection{Using NLTE analysis to derive the stellar parameters}

Finally, we used the recent\footnote{2021-01-24} version of the Potsdam Wolf-Rayet ({\scshape PoWR})\footnote{http://www.astro.physik.uni-potsdam.de/PoWR} stellar atmosphere code \citep[e.g.][]{Grafener2002, Hamann2004} to perform an analysis of the central star based on our spectroscopic data (see also \citealt{Todt2015} for details on the computing method for WR model grids). Here again, the modelling is limited by the absence of UV data -- an example of full analysis can be found in \citet{Gomez2020}. 

The main parameters of a WR model atmosphere are the luminosity $L$ and the
stellar temperature  $T_*$, which is defined at the stellar radius $R_*$ 
via the Stefan-Boltzmann law $L = 4 \pi \sigma R_*^2 T_*^4$.
In the absence of a proper Gaia
parallax measurement for the distance\footnote{We note that new distances were published based on GAIA measurements, leading to a value of 4.7$_{-1.7}^{+2.7}$ kpc \citep{Bailer2018} from the DR2 release. The latter has been recently superseded by the Gaia EDR3 release, resulting in a geometrical distance of $7.5^{+2.8}_{-2.3}\,$kpc \citep{Bailer2021} for PC 22. However the negative parallaxes and the large uncertainties of both Gaia measurements
(DR2: $-0.0235\pm0.1985$\,mas, EDR3: $-0.1045\pm0.1430$\,mas) impede the derivation of accurate distances and we will therefore not use those based on Gaia parallax measurements. %
}, we adopt a typical CS
luminosity of 6000\,$L_\odot$ together with a typical CS mass $M_*$ of 0.6\,$M_\odot$ \citep[see e.g.][]{schoenberner2005,miller-bertolami2007}, where the actual value of $M_*$ has no noticeable influence on the synthetic spectra.

Some of the parameters that  describe the stellar wind  can 
be combined in the so-called transformed radius $R_\text{t}$. This
quantity was introduced by \citet{Schmutz1989}; we define it as
\begin{equation}
\label{eq:rt}
R_{\mathrm{t}} = 
R_* \left(\frac{v_\infty}{2500 \, \mathrm{km}\,\mathrm{s^{-1}}} 
\left/
\frac{\dot M \sqrt{D}}{10^{-4} \, M_\odot \, \text{
    yr}^{-1}} \right)^{2/3}
\right. 
\end{equation}
with $v_\infty$ denoting the terminal wind velocity, $\dot M$ the
mass-loss rate, and $D$ the clumping contrast. Model spectra with equal $R_\mathrm{t}$ 
have approximately the same emission line equivalent widths, independent
of the specific combination of the particular wind parameters, as long
as $T_*$ and the chemical composition are the same.

Here we use a clumping contrast of $D=10$, as reported by \citet{Todt2008} for WC-type CSPNe, but we also tested larger values without a noticeable impact on the synthetic spectrum. Smaller values of $D$ give a worse fit.

The best fit to the observation is obtained for a transformed
radius of $\log (R_\mathrm{t}/R_\odot) = 1.27\pm0.05$. Much larger values result in a
too weak O\,{\scshape  vi}~$\lambda\lambda$~3811,3834 multiplet, while
for models with smaller value of $R_\mathrm{t}$ all lines become too
strong. 

In terms of stellar temperature, the best fit to the optical spectrum is achieved for models with
a stellar temperature of about
$T_*=130^{+3}_{-5}\,\mathrm{kK}$, which is the effective temperature at $R_*$. 
Our $R_*$ is defined at a radial Rosseland continuum optical depth $\tau_\mathrm{Ross} =20$. 
Our optimum $T_*$ corresponds to the lower limit found using 3MdB and the highest tolerance.
Temperatures below $125\,\mathrm{kK}$ lead to too strong lines of
O\,{\scshape  vi}~$\lambda\lambda$~3811,3834,
C\,{\scshape iv}~$\lambda\lambda$~5801,5812,
O\,{\scshape v}~$\lambda$~5114, and He\,{\scshape ii}~$\lambda$~4686,
while temperatures above $133\,\mathrm{kK}$ result in a too weak
O\,{\scshape vi}~$\lambda\lambda$~3811,3834 multiplet.

We note the presence of an emission feature at about 6068\,\AA, which
can be reproduced with much hotter models ($T_*>170\,\mathrm{kK}$) as
a Ne\,{\scshape viii} emission line, as reported by
\citet{Werner2007}. However, for such high temperatures the fit to the 
O\,{\scshape  vi}~$\lambda\lambda$~3811,3834 multiplet becomes much worse 
(see Fig.~\ref{fig:optical-ovi-neviii})
and also the O\,{\scshape v} lines in the model disappear. 
Unfortunately, the O\,{\scshape v} lines are even in the best fitting model
too weak to be unambiguously identified in the observation, due to its
low S/N. Otherwise their clear detection would give a more reliable
temperature estimate. 

In the absence of stellar lines with P Cygni profiles, as, e.g., typical for
the UV range, we have to estimate the terminal wind velocity from the
width of the stellar optical emission lines. In our models the
O\,{\scshape vi}~$\lambda\lambda$~3811,3834 doublet turns out to be the most sensitive line
regarding the terminal velocity. We inferred a terminal velocity of
about $v_{\infty}\approx 4500\pm500\,\mathrm{km}\,\mathrm{s}^{-1}$ with a
large uncertainty due to the low quality of the optical spectrum.
The velocity field is prescribed by a so-called $\beta$-law, where we use a value of $\beta=1$. 
We also tested different values of $\beta$, but could not achieve an improvement of the fit quality. 

The carbon abundance can be best determined from the 
C\,{\scshape iv}~$\lambda\lambda$~5801,5812 doublet. A good match to
the observation is obtained by models with a mass fraction of
$X_\mathrm{C}=0.10\pm0.02$. As there is no strong unblended helium
line in the observation, we use the He\,{\scshape ii}~$\lambda$4684
line to estimate the helium abundance and find a sufficient fit for
$X_\mathrm{He}=0.76\pm0.10$. 


As the most prominent O\,{\scshape vi} feature, the  O\,{\scshape vi}~$\lambda$ $\lambda$~3811,3834 doublet turns out to be sensitive to several parameters. We use the other, less sensitive O\,{\scshape vi} lines, $\lambda$$\lambda$ 4500,4502, $\lambda$$\lambda$ 5291,5292, and $\lambda$ 6200 (see Fig. 11) to infer an oxygen mass fraction of $X_\mathrm{O}=0.10^{+0.04}_{-0.02}$.
 

The object shows an enrichment in neon of about $X_\mathrm{Ne}=0.04^{+0.02}_{-0.01}$ by mass, as inferred from the Ne\,{\scshape vi}~$\lambda$~3902 line.
Interestingly, the model with the larger temperature (see above) gives for
the same neon abundance a good fit to the (alleged) Ne\,{\scshape viii}~$\lambda$6069 line, but
fails to reproduce the Ne\,{\scshape vi}~$\lambda$3902 line. The reason for such behaviour is unclear at the moment.

We can not detect any nitrogen line in the observation, hence we give an upper limit of $X_\mathrm{N}<0.003$, which is about four times the solar
value.

For hydrogen we can only give an upper limit for its abundance too, as
stellar Balmer lines would be blended with the corresponding
He{\,\scshape ii} lines from the Pickering series and the strong
nebular hydrogen emission lines.
We find that a hydrogen mass fraction below 25 per cent
would escape detection.

In the absence of an existing UV observation we could not
determine an iron abundance and adopted instead the solar value. 

With the distance-independent parameters fixed, we finally fitted our synthetic SED from the PoWR
model to the Gaia EDR3 photometry and our observed optical spectrum, which is only relative flux-calibrated. However, the absolute flux level of our optical spectrum was scaled to the Gaia $G_\mathrm{bp}$ magnitude. Using the reddening law from \citet{Cardelli1989} with an $R_V=3.1$ we obtained the luminosity distance of about $6.2$\,kpc and a reddening of about $E(B-V)=0.3\pm0.05$\,mag, which is slightly higher, but still consistent with the color excess inferred from $c(\mathrm{H}\beta)$.

\begin{table}
  \caption{Parameters of the CSPN of PC\,22, obtained with PoWR.}
  \label{tab:powrparameters}
\begin{tabular}{lcl}
\hline
Parameter & Value & Comment\\
\hline
$d$ [kpc]                            & $6.2\pm0.4$       & $d \propto L_\star^{1/2}$ \\
$E(B-V)$ [mag]                       & $0.30\pm0.05$     & SED fit \\
$T_\star$ [kK]                       & $130^{+3}_{-5}$   & defined at $\tau_\mathrm{Ross}=20$ \\
$\log(L_{\star}/L_\odot)$            & 3.78               &   adopted\\
$R_{\star}$ [$R_\odot$]              & $0.15^{+0.02}_{-0.01}$     &  $R_\star \propto L_\star^{1/2} $ \\ 
$\log (R_\mathrm{t}/R_\odot)$      & $1.27\pm0.05$     & transformed radius (see Eq.~(\ref{eq:rt}))\\
$\dot{M}/$M$_{\odot}$~yr$^{-1}$      & $(4.2\pm0.8)\times10^{-8}$ & $\dot{M} \propto D^{-1/2}  L^{3/4}$ \\ 
$v_{\infty}$ [km\,s$^{-1}$]          & $4500\pm500$      &  \\
$D$                                  & 10                & density contrast \\
\hline
\multicolumn{3}{l}{Chemical abundances (mass fraction)}\\
\hline
H  &  $<0.25$                   &  upper limit\\
He &  $0.76\pm0.10$             & \\
C  &  $0.10\pm0.02$             & \\
N  &  $<3\times10^{-3}$         & $4\times$ solar, upper limit \\
O  &  $0.10^{+0.04}_{-0.02}$    & \\
Ne &  $0.04^{+0.02}_{-0.01}$    & \\
Fe &  $1.4\times10^{-3}$        & solar, adopted \\
\hline
\end{tabular}
\end{table}


\begin{table*} 
\caption{Total abundances in a set of [WO1]-type PNe} 
\label{CompAbund}
\scalebox{1.0}{
\begin{tabular}{@{\extracolsep{4pt}}lccccc} \hline \hline 
Element & PC\,22$^a$ & NGC\,2371-72$^b$ & PB\,6$^c$ &  NGC\,5189$^d$ & Solar$^e$\\
\hline 
He  & 0.135                     & 0.126                 &     0.170                & 0.123                       &  0.084\\
O   & 4.79$\times$10$^{-4}$     & 1.65$\times$10$^{-4}$ &     3.47$\times$10$^{-4}$& 5.90/8.30$\times$10$^{-4}$  &  5.37$\times$10$^{-4}$\\
N   & 3.55$\times$10$^{-4}$     & 6.95$\times$10$^{-5}$ &     3.80$\times$10$^{-4}$& 4.00$\times$10$^{-4}$       &  7.24$\times$10$^{-5}$ \\
Ne  & 1.62/5.12$\times$10$^{-4}$& 9.55$\times$10$^{-5}$ &     1.00$\times$10$^{-4}$& 1.90/0.40$\times$10$^{-4}$  &  1.12$\times$10$^{-4}$\\
S   & 2.82$\times$10$^{-5}$     & 6.45$\times$10$^{-6}$ &     7.58$\times$10$^{-6}$& 1.29$\times$10$^{-5}$       & 1.45$\times$10$^{-5}$\\
Cl  & 3.31$\times$10$^{-7}$     &    ...                &     1.66$\times$10$^{-7}$& 3.71/3.47$\times$10$^{-7}$  & 1.78$\times$10$^{-7}$\\
Ar  & 4.79$\times$10$^{-6}$     & 8.00$\times$10$^{-6}$ &     8.91$\times$10$^{-7}$& 5.13/4.50$\times$10$^{-6}$  &  3.16$\times$10$^{-6}$\\
\hline
N/O  & 0.74                     & 0.42                  &      1.095               &     0.68                    & 0.13\\
Ne/O & 0.34/1.07                & 0.59                  &      0.29                &     0.32/0.068              & 0.21\\
S/O  & 0.06                     & 0.039                 &      0.022               &     0.022                   & 0.03\\
Cl/O & 0.0007                   & ...                   &      0.0005              &     0.00063/0.00059         & 0.0003\\
Ar/O & 0.01                     & 0.048                 &      0.0026              &     0.0087/0.0076           & 0.006\\
\hline
\hline
\multicolumn{6}{l}{%
  \begin{minipage}{12.5cm}%
     References: $^a$This work: Mean values obtained from the Monte Carlo method with MLA, Ne/H: values without/with Ne$^{3+}$ $^b$\citet{Gomez2020}: Average of internal regions A3 and A5, $^c$\citet{Ali2017}, $^d$\citet{GarciaRojas2013}, $^e$\citet{Lodders2010}  .%
  \end{minipage}%
}\\

\end{tabular}} 
\end{table*}

\begin{table}
  \caption{Comparison of the effective temperatures of [WO1]-type CS}
  \label{tab:Teff}
\begin{tabular}{lcl}
\hline
PN & T$_{eff}$ & References \\
\hline
 NGC 2371-72 &130 kK & \citet{Gomez2020}\\  
             & 135kK & \citet{Herald2004}\\
             &         &                    \\         
 NGC 5189    &   165kK  & \citet{Keller2014}\\
             &   135kK  & \citet{Althaus2010} \\
             &   135kK  & \citet{Koesterke2001}\\
            &         &                    \\  
 PB6        &   165kK  & \citet{Keller2014}\\
            &   110 kK & \citet{Gesicki2003}\\
            &   140kK  & \citet{Koesterke2001}\\
             &         &                    \\  
 NGC 2452  &   141kK  & \citet{Koesterke2001}\\
           &   110kK  & \citet{Stanghellini1993}\\
           &         &                    \\  
 Sand 3   &   150 kK & \citet{Keller2014}\\
          &   140kK  & \citet{Koesterke2001}\\
          &         &                    \\  
{\bf PC\,22}   & {\bf $\geq$130 kK} & {\bf This work}\\
\hline
\end{tabular}
\end{table}

\section{Discussion}\label{sec_disc} 

\subsection{A new approach for the nebular analysis of PNe}\label{sec_disc_MLA} 

In this paper we use a Monte-Carlo based method to estimate the propagation of the uncertainties due to emission line intensity measurements through the whole pipeline starting from the reddening correction, to obtain as final product the element abundances. The complete distributions of the intermediate and final parameters describing the nebula (T$_\mathrm{e}$, n$_\mathrm{e}$, ionic abundances, and total abundances) are shown in detail, leading to an easy visualization of the asymmetries of the uncertainties in some cases. The resulting distributions obtained in the previous sections depend on the initial distribution of the uncertainties on the emission line intensities, to which we added a systematic 10\% to take into account other sources of errors, for example the atomic data needed in almost any step of the pipeline. 

The final abundance uncertainties are ranging from 0.06 dex for O/H and 0.05 dex for Ar/H to $\sim$0.10 dex for N/H, Ne/H and Cl/H and finally 0.13 dex for S/H. It is practically impossible to follow every step of the process leading to these abundances, to apply the uncertainties propagation. The Monte Carlo method is the only safe way to obtain this result, nowadays being easily computable with parallel mode and multiple CPUs.

This is the first time ICFs has been determined from a grid of photoionisation models interpolated using an MLA. This allows a more versatile determination of the relation between the observed line ratio (used here instead of the ionic fraction ratios, with a proxy for the electron temperature given by the \forbr{O}{iii}{4363}{5007} line ratio) and the ICFs. We do not need here to first define the form of a function for which we will have to determine some parameters using a $\chi^2$ minimisation. This method allows also to determine the ICFs based on 6 observed line ratios, instead of one (or two in rare cases) ionic abundance ratios. 
Another advantage of using Machine Learning is the possibility to {\it tailor the ICFs calculations to the object of study}. These can then be compared to the more traditional ICFs estimations by KB94 and DIMS14. In our case, we unveil totally new ICFs for (S$^{+}$ + S$^{++}$)/O$^{++}$, Cl$^{++}$/O$^{++}$, and Ar$^{3+}$+Ar$^{4+}$ which could not have been determined otherwise. As a consequence the determination of the total abundances of Argon but also neon (with some caveat, {\it see above}) in PC\,22 are unique.

\subsection{PC\,22: a new extreme [WR]-type PN}

Since Paper I, it was clear that PC\,22 was a highly excited PN and in this work the spectroscopic analysis from the nebula and the central star confirmed our previous assertion. PC\,22 is not only a [WR]-type PN but more precisely a [WO1]-subtype, placing it in a select group of objects. Indeed, in the new edition of their CSPNe catalogue \citet{Weidmann2020} listed only 5 objects with a [WO 1]-type nucleus, namely NGC\,2371-72, NGC\,2452, PB\,6, NGC\,5189, and Sand\,3. PC\,22 is therefore a new addition to this reduced group and in the following we proceed to a comparative study with the other [WO 1] PNe.

Similarly to NGC\,2371-72 \citep{Gomez2020}, NGC\,5189 \citep{Sabin2012}, PB\,6 \citep{Dufour2015}, and NGC\,2452 \citep{Corradi1996}, PC\,22 presents a \forb{O}{iii}-dominated filamentary and clumpy morphology (often characterised as "complex") with internal (low ionisation) micro-structures under the form of \forb{N}{ii} knots. We did not find any mention of an extended optical nebula around Sand 3, but at infrared wavelengths, with WISE, a spherical or bipolar morphology (depending on $\lambda$) has been detected, as well as a single knot which displays an optical counterpart \citep{Griffith2015,Gvaramadze2020}.
When compared to the solar abundances derived by \citet{Lodders2010} (see Table \ref{CompAbund}), PC\,22 is clearly N over-abundant with (N/H)~$\approx$~4.9(N/H)$_\odot$ and (N/O)~$\approx$~5.7(N/O)$_\odot$. The same can be said about the Ne abundance if all the ionic species are used for the calculation of the ICFs, and in this case (Ne/H)~$\approx$~4.6(Ne/H)$_\odot$. 
The difference is less pronounced when it comes to He, Ar, S and Cl with variations of approximately 1.6, 1.5, 1.9 and 1.9 times their corresponding solar values. The only sub-solar abundance is that of O with (O/H)~$\approx$~0.9(O/H)$_\odot$. Among the other [WO]-type PNe for which total abundances were derived, namely NGC 2371-72 \citep{Gomez2020}, PB6 \citep{Ali2017} and NGC 5189 \citep{GarciaRojas2013}, PC\,22 shows relatively higher Ne, S and Cl chemical abundances and could therefore be considered as slightly more metal rich than its counterparts (see Table \ref{CompAbund}) . 

In terms of effective temperatures, the different methods used (3MdB, NLTE modelling, presence of \perml{Ne}{viii}{6068} line) point towards a very high T$_\mathrm{eff}$ for the CSPN of PC\,22. The NLTE PoWR analysis indicates a value of at least 130 kK. This temperature is consistent with the other known [WO1]-type CSPNe where the temperatures are in average on the order of $\sim$140 kK (see Table \ref{tab:Teff}). We also emphasize the difficulty for the model to appropriately fit the \perml{Ne}{viii}{6068} line alongside the other bright oxygen lines (mainly \perml{O}{vi}{3811,3834}).

Finally, the stellar spectral analysis suffers from the low S/N (for this purpose) and the absence of UV data. Therefore, our reported values for parameters such as the terminal velocity of the stellar wind ($v_{\infty}\approx 4500\,\mathrm{km}\,\mathrm{s}^{-1}$) and the mass-loss rate ($\dot{M}/$M$_{\odot}$~yr$^{-1}$ $\sim$ $4.2\times10^{-8}$) have to be taken carefully, the latter in particular because of the unknown distance towards PC\,22.





\section{Summary}\label{sec_con}
Following the morpho-kinematics investigation of the planetary nebula PC\,22 (Paper I), we now present a nebular and stellar analysis. 
The main findings are described in the following:
\begin{itemize}
\item
We described in detail the use of a Monte Carlo (MC) analysis combined with PyNeb to allow a sensible analysis of the photoionised nebular environment of the PN. The MC method takes into account typical uncertainties of the line intensities in the whole process, leading to abundances determination (including reddening correction, electron temperature and density, and ionic abundances). Ultimately, we find that the values of the elemental abundances are determined with $\pm$ 12\% to $\pm$ 30\% of uncertainties (0.05 to 0.13 dex). \item
We also made use of a Machine Learning technique to determine the ICFs specific to PC\,22 by employing the python implementation of XGBoost in combination with the 3MdB database. Our ICFs were also compared to those from \citet{DIMS2014} and \citet{KB1994}; and in the case of (S$^{+}$ + S$^{++}$)/O$^{++}$, Cl$^{++}$/O$^{++}$, and Ar$^{3+}$+Ar$^{4+}$ we report first time ICF estimations.
\item
The nebular analysis is based on the data from the internal region of the nebula and we obtained electronic temperatures $T_{\rm e}$ $\approx$ 10,800~K using \forb{N}{ii} and $\approx$ 13,000 K using \forb{O}{iii}. The electronic density $n_{\rm e}$ is $\approx$ 700~cm$^{-3}$. We used our ICFs to determined the total abundances and we draw a particular attention to the neon abundance where we obtained an ICF=1 after including Ne$^{3+}$ in the calculations and therefore producing variations from the estimations by DIMS14 and KB94. PC\,22 is found to be nitrogen and neon over-abundant but in the same range as its [WO]-type counterparts (see below) and also shows an oxygen deficiency (sub-solar abundances).
\item
The stellar analysis indicated a [WO1]-type classification for PC\,22 based on the presence of the so-called violet (or oxygen)- (3820\AA), blue- (4686\AA) and red- (5806\AA) bumps as well as the \perml{O}{vi}{5291,5292} and \perml{Ne}{viii}{6068} lines. The presence of the latter allowed us to constrain the stellar temperature (hence assumed to be $\geq$ 150 kK). The 3MdB was also used to reproduce the ionization state of the gas, implying effective temperatures between 140 kK and 250 kK (depending on the adopted tolerance). Finally, and in addition, the PoWR NLTE model atmosphere code was also applied to our observations and the best fit model indicated a stellar temperature of $130^{+3}_{-5}$ kK, which might be higher as the \perml{Ne}{viii}{6068} could not be properly fit. Overall, the CSPN of PC\,22 shows a high T$_\mathrm{eff}$ consistent with [WR]-type CSPNe as well as other [WO1]-type in particular. The NLTE modelling also allowed us to derived values for the high stellar wind velocity and the mass loss but these have to be taken with caution due to the low S/N of the stellar spectrum and the unknown distance.
Finally the best fit for the surface abundances (by mass fraction) of PC\,22 indicated values of 
H$<$0.25, He=0.76, C=0.10, N$<$0.003, O=0.10, Ne=0.04 and Fe=0.0014.

\end{itemize}

PC\,22 is therefore a new addition to the small group of [WO1]-type PNe, although the lack of UV spectroscopic data hampers a more complete analysis (mostly of the CSPN) as it was the case for NGC\,2371-72 \citet{Gomez2020} and NGC 6905 \citet{Gomez2022}. However, besides the presentation of new sophisticated analysis techniques, the data presented here will without a doubt contribute to the better understanding of objects with [WR]-type CSPNe.

\section*{Acknowledgments}
We thank the anonymous referee for his/her suggestions and comments which help improving the article.
LS acknowledges support from PAPIIT grant IN101819 (Mexico). 
VGL and CM acknowledge grants CONACyT / CB2015 - 254132 and UNAM / PAPIIT - IN101220. 
VMAGG acknowledges support from the Programa de Becas 
posdoctorales funded by Direcci\'{o}n General 
de Asuntos del Personal Acad\'{e}mico (DGAPA) of the Universidad 
Nacional Aut\'{o}noma de M\'{e}xico (UNAM).
MAG acknowledges support of the grant PGC2018-102184-B-I00 of the Spanish Ministerio de Ciencias, Innovaci\'on y Universidades.
Also we want to thank to the OAN-SPM staff and the CATT for time allocation 
Based on observations made with the 2.1m telescope of the Observatorio Astron\'omico Nacional at the Sierra de San Pedro M\'artir (OAN-SPM), which is a national facility operated by the Instituto de Astronom\'{\i}a of the Universidad Nacional Aut\'onoma de M\' exico.

This research has made use of the NASA/IPAC Infrared Science Archive, which is operated by the Jet Propulsion Laboratory, California Institute of Technology, under contract with the National Aeronautics and Space Administration. We have also used archival observations made with the NASA/ESA Hubble Space Telescope, and obtained from the Hubble Legacy Archive, which is a collaboration between the Space Telescope Science Institute (STScI/NASA), the Space Telescope European Coordinating Facility (ST-ECF/ESA) and the Canadian Astronomy Data Centre (CADC/NRC/CSA). 
This paper have been edited using the Overleaf facility.
\section*{Data availability}

The data underlying this article will be shared on reasonable request to the corresponding author. The python notebook used to generate most of the tables and figures, as well as the Monte Carlo computations and Machine Learning methods are available on the github repository \url{https://github.com/Morisset/PC22}


\bsp
\label{lastpage}

\end{document}

%% file: tab1_final3.tex
\,[Ne~{\sc v}]                           & 3346 & 
                  -        &           -        & 
           28.9 $\pm$  3.4 &    38.9 $\pm$  4.6 & 
           41.0 $\pm$  5.4 &    55.4 $\pm$  7.3 & 
                  -        &           -        
        \\
\,[Ne~{\sc v}]                           & 3426 & 
                  -        &           -        & 
           84.1 $\pm$  8.6 &   111.3 $\pm$ 11.4 & 
           96.5 $\pm$ 10.4 &   128.1 $\pm$ 13.8 & 
                  -        &           -        
        \\
\,[O~{\sc ii}]$\dagger$                  & 3727 & 
                  -        &           -        & 
           18.9 $\pm$  2.6 &    23.9 $\pm$  3.3 & 
           17.9 $\pm$  4.0 &    22.7 $\pm$  5.0 & 
                  -        &           -        
        \\
\ion{He}{ii}+H10                         & 3798 & 
                  -        &           -        & 
            4.5 $\pm$  2.7 &     5.7 $\pm$  3.3 & 
            4.8 $\pm$  5.2 &     6.0 $\pm$  6.5 & 
                  -        &           -        
        \\
\ion{He}{ii}+H9                          & 3835 & 
                  -        &           -        & 
            9.9 $\pm$  1.1 &    12.3 $\pm$  1.3 & 
            8.2 $\pm$  1.1 &    10.2 $\pm$  1.4 & 
                  -        &           -        
        \\
\,[Ne~{\sc iii}]                         & 3869 & 
                  -        &           -        & 
           61.7 $\pm$  6.2 &    76.3 $\pm$  7.6 & 
           72.3 $\pm$  7.3 &    89.6 $\pm$  9.0 & 
                  -        &           -        
        \\
\ion{He}{i}+H8$\dagger$                  & 3889 & 
                  -        &           -        & 
            9.8 $\pm$  1.0 &    12.1 $\pm$  1.2 & 
           10.9 $\pm$  1.2 &    13.5 $\pm$  1.4 & 
                  -        &           -        
        \\
H$\epsilon$ $\dagger$                    & 3970 & 
                  -        &           -        & 
           24.3 $\pm$  2.5 &    29.5 $\pm$  3.0 & 
           29.3 $\pm$  3.1 &    35.7 $\pm$  3.8 & 
                  -        &           -        
        \\
\,[S~{\sc ii}]                           & 4069 & 
                  -        &           -        & 
            2.1 $\pm$  0.5 &     2.5 $\pm$  0.6 & 
                  -        &           -        & 
                  -        &           -        
        \\
H$\delta$                                & 4102 & 
                  -        &           -        & 
           21.3 $\pm$  2.2 &    25.2 $\pm$  2.5 & 
           23.0 $\pm$  2.4 &    27.3 $\pm$  2.8 & 
                  -        &           -        
        \\
\ion{He}{ii}                             & 4200 & 
                  -        &           -        & 
                  -        &           -        & 
            1.7 $\pm$  0.7 &     1.9 $\pm$  0.9 & 
                  -        &           -        
        \\
H$\gamma$                                & 4341 & 
                  -        &           -        & 
           42.2 $\pm$  4.2 &    47.3 $\pm$  4.7 & 
           44.0 $\pm$  4.4 &    49.4 $\pm$  5.0 & 
                  -        &           -        
        \\
\,[O~{\sc iii}]                          & 4363 & 
                  -        &           -        & 
           12.7 $\pm$  1.3 &    14.2 $\pm$  1.4 & 
           13.2 $\pm$  1.4 &    14.7 $\pm$  1.5 & 
                  -        &           -        
        \\
\ion{He}{i}                              & 4471 & 
                  -        &           -        & 
            0.7 $\pm$  0.2 &     0.8 $\pm$  0.2 & 
            1.3 $\pm$  0.3 &     1.4 $\pm$  0.3 & 
                  -        &           -        
        \\
\ion{He}{ii}                             & 4542 & 
                  -        &           -        & 
            3.0 $\pm$  0.4 &     3.2 $\pm$  0.4 & 
            3.5 $\pm$  0.6 &     3.7 $\pm$  0.7 & 
                  -        &           -        
        \\
\ion{N}{iii}                             & 4634 & 
                  -        &           -        & 
            1.0 $\pm$  0.2 &     1.0 $\pm$  0.2 & 
                  -        &           -        & 
                  -        &           -        
        \\
\ion{N}{iii}+\ion{O}{ii}                 & 4640 & 
                  -        &           -        & 
            2.3 $\pm$  0.3 &     2.4 $\pm$  0.3 & 
            2.2 $\pm$  0.4 &     2.3 $\pm$  0.5 & 
                  -        &           -        
        \\
\ion{He}{ii}                             & 4686 & 
          112.8 $\pm$ 13.9 &   120.9 $\pm$ 14.9 & 
          120.8 $\pm$ 12.1 &   125.3 $\pm$ 12.5 & 
          120.8 $\pm$ 12.1 &   125.3 $\pm$ 12.5 & 
          158.9 $\pm$ 22.4 &   160.8 $\pm$ 22.7 
        \\
\ion{He}{i}+[Ar~{\sc iv}]                & 4711 & 
                  -        &           -        & 
           17.9 $\pm$  1.8 &    18.5 $\pm$  1.9 & 
           18.6 $\pm$  1.9 &    19.1 $\pm$  1.9 & 
                  -        &           -        
        \\
\,[Ne~{\sc iv}]                          & 4726 & 
                  -        &           -        & 
            1.9 $\pm$  0.2 &     1.9 $\pm$  0.2 & 
            2.5 $\pm$  0.3 &     2.6 $\pm$  0.3 & 
                  -        &           -        
        \\
\,[Ar~{\sc iv}]                          & 4740 & 
                  -        &           -        & 
           13.2 $\pm$  1.3 &    13.5 $\pm$  1.4 & 
           14.1 $\pm$  1.4 &    14.5 $\pm$  1.5 & 
                  -        &           -        
        \\
H$\beta$                                 & 4861 & 
          100.0 $\pm$ 0.0 &   100.0 $\pm$ 0.0 & 
          100.0 $\pm$ 0.0 &   100.0 $\pm$ 0.0 & 
          100.0 $\pm$ 0.0 &   100.0 $\pm$ 0.0 & 
          100.0 $\pm$ 0.0 &   100.0 $\pm$ 0.0 
        \\
\,[O~{\sc iii}]                          & 4959 & 
          309.3 $\pm$ 31.0 &   298.4 $\pm$ 29.9 & 
          343.9 $\pm$ 34.4 &   337.5 $\pm$ 33.8 & 
          362.3 $\pm$ 36.2 &   355.4 $\pm$ 35.5 & 
          431.0 $\pm$ 43.4 &   428.4 $\pm$ 43.1 
        \\
\,[O~{\sc iii}]                          & 5007 & 
         1070.5 $\pm$107.2 &  1015.7 $\pm$101.7 & 
         1027.0 $\pm$102.7 &   999.1 $\pm$ 99.9 & 
         1066.1 $\pm$106.6 &  1037.0 $\pm$103.7 & 
         1551.9 $\pm$155.6 &  1538.1 $\pm$154.2 
        \\
\ion{He}{ii}                             & 5411 & 
                  -        &           -        & 
           10.0 $\pm$  1.0 &     9.1 $\pm$  0.9 & 
           10.4 $\pm$  1.1 &     9.5 $\pm$  1.0 & 
                  -        &           -        
        \\
\,[Cl~{\sc iii}]                         & 5518 & 
                  -        &           -        & 
            1.5 $\pm$  0.2 &     1.4 $\pm$  0.1 & 
            1.8 $\pm$  0.2 &     1.6 $\pm$  0.2 & 
                  -        &           -        
        \\
\,[Cl~{\sc iii}]                         & 5538 & 
                  -        &           -        & 
            1.2 $\pm$  0.2 &     1.1 $\pm$  0.1 & 
            1.2 $\pm$  0.2 &     1.1 $\pm$  0.2 & 
                  -        &           -        
        \\
\,[N~{\sc ii}]                           & 5755 & 
                  -        &           -        & 
            0.8 $\pm$  0.1 &     0.7 $\pm$  0.1 & 
            0.8 $\pm$  0.2 &     0.7 $\pm$  0.2 & 
                  -        &           -        
        \\
\ion{He}{i}                              & 5876 & 
                  -        &           -        & 
            4.9 $\pm$  0.5 &     4.3 $\pm$  0.4 & 
            6.0 $\pm$  0.6 &     5.1 $\pm$  0.6 & 
                  -        &           -        
        \\
\ion{He}{ii}                             & 6235 & 
                  -        &           -        & 
            0.5 $\pm$  0.1 &     0.4 $\pm$  0.1 & 
            0.8 $\pm$  0.2 &     0.6 $\pm$  0.2 & 
                  -        &           -        
        \\
\,[O~{\sc i}]                            & 6300 & 
                  -        &           -        & 
            1.0 $\pm$  0.1 &     0.8 $\pm$  0.1 & 
            0.6 $\pm$  0.1 &     0.5 $\pm$  0.1 & 
                  -        &           -        
        \\
\,[S~{\sc iii}]+\ion{He}{ii}             & 6312 & 
                  -        &           -        & 
            7.9 $\pm$  0.8 &     6.5 $\pm$  0.7 & 
            9.1 $\pm$  0.9 &     7.5 $\pm$  0.8 & 
                  -        &           -        
        \\
\ion{He}{ii}                             & 6407 & 
                  -        &           -        & 
            0.6 $\pm$  0.2 &     0.5 $\pm$  0.1 & 
                  -        &           -        & 
                  -        &           -        
        \\
\,[Ar~{\sc V}]                           & 6435 & 
                  -        &           -        & 
            3.5 $\pm$  0.4 &     2.8 $\pm$  0.3 & 
            3.9 $\pm$  0.4 &     3.2 $\pm$  0.3 & 
                  -        &           -        
        \\
\ion{He}{ii}                             & 6527 & 
                  -        &           -        & 
            0.6 $\pm$  0.2 &     0.5 $\pm$  0.2 & 
            1.0 $\pm$  0.4 &     0.8 $\pm$  0.4 & 
                  -        &           -        
        \\
\,[N~{\sc ii}]                           & 6548 & 
                  -        &           -        & 
           15.2 $\pm$  1.5 &    12.3 $\pm$  1.2 & 
           18.9 $\pm$  1.9 &    15.2 $\pm$  1.5 & 
                  -        &           -        
        \\
H$\alpha$                                & 6563 & 
          433.5 $\pm$ 44.3 &   286.3 $\pm$ 29.2 & 
          348.7 $\pm$ 34.9 &   280.7 $\pm$ 28.1 & 
          421.8 $\pm$ 42.2 &   338.8 $\pm$ 33.9 & 
          307.3 $\pm$ 35.4 &   286.3 $\pm$ 33.0 
        \\
\,[N~{\sc ii}]                           & 6584 & 
                  -        &           -        & 
           48.4 $\pm$  4.8 &    38.9 $\pm$  3.9 & 
           53.5 $\pm$  5.4 &    42.9 $\pm$  4.3 & 
           51.1 $\pm$ 11.3 &    47.5 $\pm$ 10.5 
        \\
\ion{He}{i}$\dagger$                     & 6678 & 
                  -        &           -        & 
                  -        &           -        & 
            1.7 $\pm$  0.6 &     1.3 $\pm$  0.5 & 
                  -        &           -        
        \\
\,[S~{\sc ii}]                           & 6716 & 
                  -        &           -        & 
            9.6 $\pm$  1.0 &     7.6 $\pm$  0.8 & 
           11.0 $\pm$  1.1 &     8.7 $\pm$  0.9 & 
                  -        &           -        
        \\
\,[S~{\sc ii}]                           & 6731 & 
                  -        &           -        & 
            9.1 $\pm$  0.9 &     7.2 $\pm$  0.7 & 
           10.7 $\pm$  1.1 &     8.4 $\pm$  0.9 & 
                  -        &           -        
        \\
\hline 
c(H$\beta$) & & & 0.61$^{*}$ & & 0.32 & & 0.32 & & 0.10$^{*}$ \\
\hline 
log H$\beta$ (erg/s/m$^{2}$) &&-15.35& -14.74 & -13.78& -13.46 &-14.07 & -13.75 &-15.64 & -15.54 \\

%% file: tab2_final3.tex
\,[Ne~{\sc v}]                      & 3345 &    44.08 $\pm$  5.44 \\
\,[Ne~{\sc v}]                      & 3426 &   114.56 $\pm$ 11.97 \\
\,[O~{\sc ii}]$\dagger$             & 3727 &    25.72 $\pm$  3.97 \\
\ion{He}{ii}+H10                    & 3798 &     5.59 $\pm$  4.43 \\
\ion{He}{ii}+H9                     & 3835 &     9.55 $\pm$  1.14 \\
\,[Ne~{\sc iii}]                    & 3869 &    80.06 $\pm$  8.03 \\
\ion{He}{i}+H8$\dagger$             & 3889 &    12.82 $\pm$  1.32 \\
H$\epsilon$ $\dagger$               & 3969 &    36.94 $\pm$  3.80 \\
\,[S~{\sc ii}]                      & 4070 &     2.04 $\pm$  0.77 \\
H$\delta$                           & 4102 &    26.65 $\pm$  2.71 \\
\ion{He}{ii}                        & 4199 &     1.89 $\pm$  0.61 \\
H$\gamma$                           & 4341 &    48.40 $\pm$  4.85 \\
\,[O~{\sc iii}]                     & 4363 &    14.54 $\pm$  1.48 \\
\ion{He}{i}                         & 4472 &     0.93 $\pm$  0.22 \\
\ion{He}{ii}                        & 4542 &     3.63 $\pm$  0.54 \\
\ion{N}{iii}                        & 4633 &     0.61 $\pm$  0.21 \\
\ion{N}{iii}+\ion{O}{ii}            & 4642 &     1.27 $\pm$  0.30 \\
\ion{He}{ii}                        & 4686 &   127.87 $\pm$ 12.79 \\
\ion{He}{i}+[Ar~{\sc iv}]           & 4712 &    18.95 $\pm$  1.91 \\
\,[Ne~{\sc iv}]                     & 4725 &     2.19 $\pm$  0.27 \\
\,[Ar~{\sc iv}]                     & 4740 &    14.37 $\pm$  1.45 \\
H$\beta$                            & 4862 &   100.00 $\pm$  0.00 \\
\,[O~{\sc iii}]                     & 4960 &   352.26 $\pm$ 35.23 \\
\,[O~{\sc iii}]                     & 5007 &  1050.15 $\pm$105.02 \\
\ion{He}{ii}                        & 5412 &     9.63 $\pm$  0.97 \\
\,[Cl~{\sc iii}]                    & 5519 &     1.49 $\pm$  0.16 \\
\,[Cl~{\sc iii}]                    & 5538 &     1.15 $\pm$  0.18 \\
\,[N~{\sc ii}]                      & 5755 &     0.77 $\pm$  0.13 \\
\ion{He}{i}                         & 5876 &     4.79 $\pm$  0.50 \\
\ion{He}{ii}                        & 6236 &     0.52 $\pm$  0.12 \\
\,[O~{\sc i}]                       & 6301 &     0.80 $\pm$  0.11 \\
\,[S~{\sc iii}]+\ion{He}{ii}        & 6312 &     7.20 $\pm$  0.73 \\
\ion{He}{ii}                        & 6407 &     0.47 $\pm$  0.17 \\
\,[Ar~{\sc V}]                      & 6436 &     3.02 $\pm$  0.31 \\
\ion{He}{ii}                        & 6527 &     0.66 $\pm$  0.26 \\
\,[N~{\sc ii}]                      & 6549 &    14.53 $\pm$  1.46 \\
H$\alpha$                           & 6563 &   316.28 $\pm$ 31.63 \\
\,[N~{\sc ii}]                      & 6584 &    41.52 $\pm$  4.15 \\
\ion{He}{i}$\dagger$                & 6680 &     2.35 $\pm$  0.41 \\
\,[S~{\sc ii}]                      & 6716 &     8.58 $\pm$  0.86 \\
\,[S~{\sc ii}]                      & 6731 &     8.17 $\pm$  0.85 \\
\hline 
c(H$\beta$) & & {\bf <0.35$\pm$0.13>}\\
\hline

%% file: table_tene.tex
T$_\mathrm{e}$\,[N~{\sc ii}] (K) with n$_\mathrm{e}$\,[S~{\sc ii}]  & 10802 & 10936 & 10907 & 976 \\ 
T$_\mathrm{e}$\,[O~{\sc iii}] (K) with n$_\mathrm{e}$\,[Cl~{\sc iii}] & 13024 & 13205 & 13134 & 910 \\ 
T$_\mathrm{e}$\,[O~{\sc iii}] (K) with n$_\mathrm{e}$\,[Ar~{\sc iv}] & 12849 & 13674 & 13258 & 1488 \\ 
log n$_\mathrm{e}$\,[S~{\sc ii}] (cm$^{-3}$) with T$_\mathrm{e}$\,[N~{\sc ii}] & 2.78 & 2.75 & 2.79 & 0.3 \\ 
log n$_\mathrm{e}$\,[Cl~{\sc iii}] (cm$^{-3}$) with T$_\mathrm{e}$\,[O~{\sc iii}] & 2.73 & 2.89 & 2.97 & 0.4 \\ 
log n$_\mathrm{e}$\,[Ar~{\sc iv}] (cm$^{-3}$) with T$_\mathrm{e}$\,[O~{\sc iii}] & 2.76 & 2.97 & 3.05 & 0.4 \\ 
\hline
Adopted values: &&&&\\
Low ionisation region &T$_\mathrm{e}$=10936 K &&log n$_\mathrm{e}$=2.75 cm$^{-3}$&\\
High ionisation region &T$_\mathrm{e}$=13440 K &&log n$_\mathrm{e}$= 2.93 cm$^{-3}$&\\

%% file: table_ionic.tex
log Ar$^{+3}$/H$^+$ & -5.67 &  -5.70 & -5.70 & 0.09 \\ 
log Ar$^{+4}$/H$^+$ & -6.08 &  -6.13 & -6.13 & 0.10 \\ 
log Cl$^{++}$/H$^+$ & -6.88 &  -6.90 & -6.90 & 0.12 \\ 
log He$^+$/H$^+$ & -1.43 &  -1.45 & -1.45 & 0.05 \\ 
log He$^{++}$/H$^+$ & -1.00 &  -1.01 & -1.00 & 0.05 \\ 
log N$^+$/H$^+$ & -5.15 &  -5.18 & -5.18 & 0.12 \\ 
log Ne$^{++}$/H$^+$ & -4.47 &  -4.49 & -4.49 & 0.10 \\ 
log Ne$^{+3}$/H$^+$ & -3.65 &  -3.75 & -3.74 & 0.21 \\ 
log Ne$^{+4}$/H$^+$ & -4.29 &  -4.31 & -4.31 & 0.11 \\ 
log O$^0$/H$^+$ & -5.94 &  -5.97 & -5.98 & 0.15 \\ 
log O$^+$/H$^+$ & -5.13 &  -5.11 & -5.12 & 0.17 \\ 
log O$^{++}$/H$^+$ & -3.78 &  -3.83 & -3.82 & 0.10 \\ 
log S$^{+}$/H$^+$ & -6.29 &  -6.31 & -6.31 & 0.15 \\ 
log S$^{++}$/H$^+$ & -4.93 &  -4.96 & -4.97 & 0.17 \\ 

%% file: table_icf.tex
log ICF(O$^+$ + O$^{++}$) & 0.46 & 0.45 & 0.45 & 0.04 & ML 1 \\ 
log ICF(O$^+$ + O$^{++}$) & 0.38 & 0.38 & 0.38 & 0.04 & KB94 1 \\ 
log ICF(O$^+$ + O$^{++}$) & 0.43 & 0.44 & 0.44 & 0.05 & DIMS14 1 \\ 
log ICF(N$^+$/O$^+$) & -0.13 & -0.12 & -0.13 & 0.04 & ML 2 \\ 
log ICF(N$^+$/O$^+$) & 0.00 & -0.00 & 0.00 & 0.00 & KB94 2 \\ 
log ICF(N$^+$/O$^+$) & -0.13 & -0.13 & -0.13 & 0.00 & DIMS14 2 \\ 
log ICF(Ne$^{++}$/O$^{++}$) & 0.02 & 0.03 & 0.03 & 0.02 & ML 3 \\ 
log ICF(Ne$^{++}$/O$^{++}$) & -0.00 & -0.00 & 0.00 & 0.00 & KB94 3 \\ 
log ICF(Ne$^{++}$/O$^{++}$) & 0.08 & 0.08 & 0.08 & 0.03 & DIMS14 3 \\ 
log ICF(Ne$^{+2}$ + Ne$^{+3}$ + Ne$^{+4}$) & 0.00 & 0.00 & 0.00 & 0.00 & ML234 4 \\ 
log ICF(Ne$^{+2}$ + Ne$^{+4}$) & 0.29 & 0.29 & 0.29 & 0.02 & ML24 4 \\ 
log ICF(Ne$^{+2}$ + Ne$^{+4}$) & 0.18 & 0.18 & 0.18 & 0.00 & KB94 4 \\ 
log ICF(Ne$^{+2}$ + Ne$^{+4}$) & 0.23 & 0.23 & 0.23 & 0.01 & DIMS14 4 \\ 
log ICF(S$^+$ + S$^{++}$) & 0.35 & 0.38 & 0.37 & 0.05 & ML 5 \\ 
log ICF(S$^+$ + S$^{++}$) & 0.42 & 0.41 & 0.41 & 0.06 & KB94 5 \\ 
log ICF((S$^+$ + S$^{++}$)/O$^{++}$) & -0.05 & -0.05 & -0.05 & 0.03 & ML 6 \\ 
log ICF((S$^+$ + S$^{++}$)/O$^{+}$) & -1.49 & -1.52 & -1.52 & 0.10 & ML 7 \\ 
log ICF((S$^+$ + S$^{++}$)/O$^{+}$) & -1.08 & -1.04 & -1.04 & 0.14 & DIMS14 7 \\ 
log ICF(Cl$^{++}$/O$^{+}$) & -1.39 & -1.40 & -1.40 & 0.09 & ML 9 \\ 
log ICF(Cl$^{++}$/O$^{+}$) & -1.06 & -1.02 & -1.02 & 0.15 & DIMS14 9 \\ 
log ICF(Cl$^{++}$/O$^{++}$) & 0.08 & 0.09 & 0.09 & 0.02 & ML 8 \\ 
log ICF(Ar$^{+3}$+ Ar$^{+4}$) & 0.22 & 0.21 & 0.21 & 0.02 & ML 10 \\ 

%% file: table_elem.tex
log He/H & -0.87 & -0.87 & -0.87 & 0.04 & ICF = 1 \\ 
log O/H & -3.31 & -3.35 & -3.34 & 0.10 & ML 1 \\ 
log O/H & -3.39 & -3.42 & -3.41 & 0.10 & KB94 1 \\ 
log O/H & -3.34 & -3.37 & -3.36 & 0.10 & DIMS14 1 \\ 
log N/H & -3.46 & -3.53 & -3.54 & 0.17 & ML 2 \\ 
log N/H & -3.41 & -3.48 & -3.49 & 0.16 & KB94 2 \\ 
log N/H & -3.49 & -3.56 & -3.56 & 0.15 & DIMS14 2 \\ 
log Ne/H & -3.98 & -3.99 & -3.98 & 0.10 & ML 3 \\ 
log Ne/H & -3.50 & -3.58 & -3.57 & 0.17 & ML234 4 \\ 
log Ne/H & -3.78 & -3.80 & -3.80 & 0.11 & ML24 4 \\ 
log Ne/H & -4.08 & -4.09 & -4.08 & 0.10 & KB94 3 \\ 
log Ne/H & -3.89 & -3.92 & -3.91 & 0.10 & KB94 4 \\ 
log Ne/H & -4.00 & -4.00 & -4.00 & 0.12 & DIMS14 3 \\ 
log Ne/H & -3.84 & -3.86 & -3.86 & 0.10 & DIMS14 4 \\ 
log S/H & -4.56 & -4.56 & -4.57 & 0.17 & ML 5 \\ 
log S/H & -4.49 & -4.51 & -4.52 & 0.18 & ML 6 \\ 
log S/H & -4.49 & -4.53 & -4.54 & 0.14 & KB94 5 \\ 
log S/H & -4.20 & -4.24 & -4.24 & 0.14 & DIMS14 6 \\ 
log S/H & -4.58 & -4.69 & -4.70 & 0.16 & ML 7 \\ 
log Cl/H & -6.33 & -6.33 & -6.33 & 0.14 & ML 8 \\ 
log Cl/H & -6.15 & -6.17 & -6.17 & 0.10 & DIMS14 8 \\ 
log Cl/H & -6.45 & -6.53 & -6.53 & 0.13 & ML 9 \\ 
log Ar/H & -5.30 & -5.35 & -5.35 & 0.09 & ML 10 \\ 

%% file: main_v3.bbl
\begin{thebibliography}{99}

\bibitem[Acker \& Neiner(2003)]{Acker2003} Acker, A., \& Neiner,
  C.\ 2003, \aap, 403, 659
 
 \bibitem[\protect\citeauthoryear{Ali \& Dopita}{2017}]{Ali2017} Ali A., Dopita M.~A., 2017, PASA, 34, e036

\bibitem[\protect\citeauthoryear{Althaus et al.}{2010}]{Althaus2010} Althaus L.~G., C{\'o}rsico A.~H., Isern J., Garc{\'\i}a-Berro E., 2010, A\&ARv, 18, 471. doi:10.1007/s00159-010-0033-1


\bibitem[\protect\citeauthoryear{Bailer-Jones et al.}{2018}]{Bailer2018} Bailer-Jones C.~A.~L., Rybizki J., Fouesneau M., Mantelet G., Andrae R., 2018, AJ, 156, 58. doi:10.3847/1538-3881/aacb21

\bibitem[\protect\citeauthoryear{Bailer-Jones et al.}{2021}]{Bailer2021} Bailer-Jones C.~A.~L., Rybizki J., Fouesneau M., Demleitner M., Andrae R., 2021, AJ, 161, 147. doi:10.3847/1538-3881/abd806


 \bibitem[\protect\citeauthoryear{Brocklehurst}{1971}] {Brocklehurst1971} Brocklehurst M., 1971, MNRAS, 153, 471 
  
\bibitem[\protect\citeauthoryear{Butler, K. \& Zeippen, C. J.} {1989}]{1989Butler_aap208} Butler, K. \& Zeippen, C. J. , 1989, A\&A, 208,  337-344
\bibitem[Cardelli et al. (1989)]{Cardelli1989} Cardelli, J.~A.~A., Clayton, G.~C., Mathis, J.~S.\ 1989, \apj, 345, 245

\bibitem[\protect\citeauthoryear{Chen \& Guestrin}{2016}]{Chen2016}Chen T., Guestrin C., 2016, Proceedings of the 22nd ACM SIGKDD International Conference on Knowledge Discovery and Data Mining, 785--794

\bibitem[\protect\citeauthoryear{Corradi et al.}{1996}]{Corradi1996} Corradi R.~L.~M., Manso R., Mampaso A., Schwarz H.~E., 1996, Astronomy and Astrophysics, 313, 913


\bibitem[Crowther et al.(1998)]{Crowther1998} Crowther, P.~A., De
Marco, O., \& Barlow, M.~J.\ 1998, \mnras, 296, 367

\bibitem[\protect\citeauthoryear{Cuesta, Phillips \& Mampaso}{1993}]{Cuesta1993} Cuesta L., Phillips J.~P., Mampaso A., 1993, A\&A, 267, 199

\bibitem[\protect\citeauthoryear{Dance et al.}{2013}]{2013Dance_mnra} Dance M., Palay E., Nahar S.~N., Pradhan A.~K., 2013, MNRAS, 435, 1576. doi:10.1093/mnras/stt1398



\bibitem[\protect\citeauthoryear{Delgado-Inglada, Morisset, \& Stasi{\'n}ska}{2014}]{DIMS2014} Delgado-Inglada G., Morisset C., Stasi{\'n}ska G., 2014, MNRAS, 440, 536

\bibitem[\protect\citeauthoryear{Dufour et al.}{2015}]{Dufour2015} Dufour R.~J., Kwitter K.~B., Shaw R.~A., Henry R.~B.~C., Balick B., Corradi R.~L.~M., 2015, ApJ, 803, 23

\bibitem[\protect\citeauthoryear{Frew, Parker, \& Boji{\v{c}}i{\'c}}{2016}]{Frew2016} Frew D.~J., Parker Q.~A., Boji{\v{c}}i{\'c} I.~S., 2016, MNRAS, 455, 1459. doi:10.1093/mnras/stv1516

\bibitem[\protect\citeauthoryear{Froese Fischer, C. \& Tachiev, G.} {2004}]{2004Froese-Fischer_Atom87} Froese Fischer, C. \& Tachiev, G. , 2004, Atomic Data and Nuclear Data Tables, 87,  1-184

\bibitem[\protect\citeauthoryear{Froese Fischer, C., Tachiev, G., \& Irimia, A.} {2006}]{2006Froese-Fischer_Atom92} Froese Fischer, C., Tachiev, G., \& Irimia, A. , 2006, Atomic Data and Nuclear Data Tables, 92,  607-81

\bibitem[\protect\citeauthoryear{Galavis, M. E., Mendoza, C., \& Zeippen, C. J.} {1995}]{1995Galavis_aaps111} Galavis, M. E., Mendoza, C., \& Zeippen, C. J. , 1995, A\&A, 111,  347

\bibitem[\protect\citeauthoryear{Galavis, M. E., Mendoza, C., \& Zeippen, C. J.} {1997}]{1997Galavis_aaps123} Galavis, M. E., Mendoza, C., \& Zeippen, C. J. , 1997, A\&A, 123,  159-171


\bibitem[\protect\citeauthoryear{Garc{\'\i}a-Rojas et al.}{2013}]{GarciaRojas2013} Garc{\'\i}a-Rojas J., Pe{\~n}a M., Morisset C., Delgado-Inglada G., Mesa-Delgado A., Ruiz M.~T., 2013, A\&A, 558, A122


\bibitem[\protect\citeauthoryear{Gesicki, Acker, \& Zijlstra}{2003}]{Gesicki2003} Gesicki K., Acker A., Zijlstra A.~A., 2003, A\&A, 400, 957. doi:10.1051/0004-6361:20030079

\bibitem[\protect\citeauthoryear{Giles, K.} {1981}]{1981Giles_mnra195} Giles, K. , 1981, MNRAS, 195,  63P-66P

\bibitem[\protect\citeauthoryear{Godefroid, M. \& Fischer, C. F.} {1984}]{1984Godefroid_Jour17} Godefroid, M. \& Fischer, C. F. , 1984, Journal of Physics B Atomic Molecular Physics, 17,  681-692

\bibitem[G{\'o}mez-Gonz{\'a}lez et al.(2020)]{GomezGonzalez2020} G{\'o}mez-Gonz{\'a}lez, V.~M.~A. et al.\ 2020, MNRAS, 493, 3879

\bibitem[\protect\citeauthoryear{G{\'o}mez-Gonz{\'a}lez et al.}{2020}]{Gomez2020} G{\'o}mez-Gonz{\'a}lez V.~M.~A. et al., 2020, MNRAS, 496, 959

\bibitem[\protect\citeauthoryear{G{\'o}mez-Gonz{\'a}lez et al.}{2022}]{Gomez2022} G{\'o}mez-Gonz{\'a}lez V.~M.~A., Rubio G., Toal{\'a} J.~A., Guerrero M.~A., Sabin L., Todt H., G{\'o}mez-Llanos V., et al., 2022, MNRAS, 509, 974. doi:10.1093/mnras/stab3042



\bibitem[\protect\citeauthoryear{Gr{\"a}fener, Koesterke, \& Hamann}{2002}]{Grafener2002} Gr{\"a}fener G., Koesterke L., Hamann W.-R., 2002, A\&A, 387, 244. doi:10.1051/0004-6361:20020269


\bibitem[\protect\citeauthoryear{Griffith et al.}{2015}]{Griffith2015} Griffith R.~L., Wright J.~T., Maldonado J., Povich M.~S., Sigur{\dj}sson S., Mullan B., 2015, ApJS, 217, 25. doi:10.1088/0067-0049/217/2/25


\bibitem[\protect\citeauthoryear{Guill{\'e}n, et al.}{2013}]{Guillen2013} Guill{\'e}n P.~F., V{\'a}zquez R., Miranda L.~F., Zavala S., Contreras M.~E., Ayala S., Ortiz-Ambriz A., 2013, MNRAS, 432, 2676


\bibitem[\protect\citeauthoryear{Gvaramadze et al.}{2020}]{Gvaramadze2020} Gvaramadze V.~V., Kniazev A.~Y., Gr{\"a}fener G., Langer N., 2020, MNRAS, 492, 3316. doi:10.1093/mnras/stz3639


\bibitem[\protect\citeauthoryear{Hamann \& Gr{\"a}fener}{2004}]{Hamann2004} Hamann W.-R., Gr{\"a}fener G., 2004, A\&A, 427, 697. doi:10.1051/0004-6361:20040506

\bibitem[\protect\citeauthoryear{Herald \& Bianchi}{2004}]{Herald2004} Herald J.~E., Bianchi L., 2004, ApJ, 609, 378. doi:10.1086/421010



\bibitem[\protect\citeauthoryear{Keller, Bianchi, \& Maciel}{2014}]{Keller2014} Keller G.~R., Bianchi L., Maciel W.~J., 2014, MNRAS, 442, 1379. doi:10.1093/mnras/stu878


\bibitem[\protect\citeauthoryear{Kingsburgh \& Barlow}{1994}]{KB1994} Kingsburgh R.~L., Barlow M.~J., 1994, MNRAS, 271, 257

\bibitem[Kingsburgh(1995)]{Kingsburgh1995} Kingsburgh R.~L., Barlow M.~J.,\& Storey P.~J.\ 1995, \aap, 295, 75

\bibitem[\protect\citeauthoryear{Kisielius, R., Storey, P. J., Ferland, G. J., \& Keenan, F. P.} {2009}]{2009Kisielius_mnra397} Kisielius, R., Storey, P. J., Ferland, G. J., \& Keenan, F. P. , 2009, MNRAS, 397,  903-912

\bibitem[\protect\citeauthoryear{Koesterke}{2001}]{Koesterke2001} Koesterke L., 2001, Ap\&SS, 275, 41


\bibitem[\protect\citeauthoryear{Lodders}{2010}]{Lodders2010} Lodders K., 2010, ASSP, 16, 379

\bibitem[\protect\citeauthoryear{Luridiana, Morisset \& Shaw}{2015}]{Luridiana2015} Luridiana V., Morisset C., Shaw R.~A., 2015, A\&A, 573, A42

\bibitem[\protect\citeauthoryear{McLaughlin, B. M. \& Bell, K. L.} {2000}]{2000McLaughlin_Jour33} McLaughlin, B. M. \& Bell, K. L. , 2000, Journal of Physics B Atomic Molecular Physics, 33,  597-613

\bibitem[\protect\citeauthoryear{Mendoza, C. \& Zeippen, C. J.} {1982}]{1982Mendoza_mnra199} Mendoza, C. \& Zeippen, C. J. , 1982, MNRAS, 199,  1025-1032



\bibitem[{{Miller Bertolami} \& {Althaus}(2007)}]{miller-bertolami2007} {Miller Bertolami}, M.~M. \& {Althaus}, L.~G. 2007, \mnras, 380, 763



\bibitem[\protect\citeauthoryear{Morisset, Delgado-Inglada \& Flores-Fajardo}{2015}]{Morisset2015} Morisset C., Delgado-Inglada G., Flores-Fajardo N., 2015, RMxAA, 51, 103

\bibitem[Osterbrock \& Ferland (2006)]{Osterbrock2006} Osterbrock, D.~E.,
 \& Ferland, G.~J.\ 2006, Astrophysics of gaseous nebulae and active galactic nuclei


\bibitem[\protect\citeauthoryear{Peimbert \& Costero}{1969}]{Peimbert1969} Peimbert M., Costero R., 1969, Boletin de los Observatorios Tonantzintla y Tacubaya, 5, 3




\bibitem[Press et al.(1992)]{Press1992} Press, W.~H., Teukolsky, S.~A., Vetterling, W.~T., et al.\ 1992, Cambridge: University Press

\bibitem[\protect\citeauthoryear{Porter, R. L., Ferland, G. J., Storey, P. J., \& Detisch, M. J.} {2012, corrected 2013}]{2012Porter_mnra425} Porter, R. L., Ferland, G. J., Storey, P. J., \& Detisch, M. J. , 2012, MNRAS, 425,  L28-L31

\bibitem[\protect\citeauthoryear{Porter, R. L., Ferland, G. J., Storey, P. J., & Detisch, M. J.} {2013}]{2013Porter_mnra433} Porter, R. L., Ferland, G. J., Storey, P. J., \& Detisch, M. J. , 2013, MNRAS, 433,  L89-L90



\bibitem[\protect\citeauthoryear{Ramsbottom, C. A. \& Bell, K. L.} {1997}]{1997Ramsbottom_Atom66} Ramsbottom, C. A. \& Bell, K. L. , 1997, Atomic Data and Nuclear Data Tables, 66,  65

\bibitem[\protect\citeauthoryear{Rubio, et al.}{2015}]{Rubio2015} Rubio G., V{\'a}zquez R., Ramos-Larios G., Guerrero M.~A., Olgu{\'\i}n L., Guill{\'e}n P.~F., Mata H., 2015, MNRAS, 446, 1931

\bibitem[\protect\citeauthoryear{Rynkun, P., Gaigalas, G., \& Jönsson, P.} {2019}]{2019Rynkun_aap623} Rynkun, P., Gaigalas, G., \& Jönsson, P. , 2019, A\&A, 623,  A155


\bibitem[\protect\citeauthoryear{Sabin et al.}{2012}]{Sabin2012} Sabin L., V{\'a}zquez R., L{\'o}pez J.~A., Garc{\'\i}a-D{\'\i}az M.~T., Ramos-Larios G., 2012, RMxAA, 48, 165


\bibitem[\protect\citeauthoryear{Sabin, et al.}{2017}]{Sabin2017} Sabin L., et al., 2017, MNRAS, 467, 3056

\bibitem[{{Schmutz} {et~al.}(1989){Schmutz}, {Hamann}, \&  {Wessolowski}}]{Schmutz1989}
{Schmutz}, W., {Hamann}, W.-R., \& {Wessolowski}, U. 1989, \aap, 210, 23

\bibitem[{{Sch{\"o}nberner} {et~al.}(2005){Sch{\"o}nberner},
  {Jacob}, {Steffen}, {Perinotto}, {Corradi}, \& {Acker}}]{schoenberner2005}
{Sch{\"o}nberner}, D., {Jacob}, R., {Steffen}, M., {et~al.} 2005,
  \aap, 431, 963


\bibitem[\protect\citeauthoryear{Stanghellini, Corradi, \& Schwarz}{1993}]{Stanghellini1993} Stanghellini L., Corradi R.~L.~M., Schwarz H.~E., 1993, A\&A, 279, 521

\bibitem[\protect\citeauthoryear{Storey, P. J. \& Hummer, D. G.} {1995}]{1995Storey_mnra272} Storey, P. J. \& Hummer, D. G. , 1995, MNRAS, 272,  41--48


\bibitem[\protect\citeauthoryear{Storey, P. J., Sochi, T., \& Badnell, N. R.} {2014}]{2014Storey_mnra441} Storey, P. J., Sochi, T., \& Badnell, N. R. , 2014, MNRAS, 441,  3028-3039

\bibitem[\protect\citeauthoryear{Tajitsu \& Tamura}{1998}]{Tajitsu1998} Tajitsu A., Tamura S., 1998, AJ, 115, 1989. doi:10.1086/300315

\bibitem[\protect\citeauthoryear{Tayal, S. S. \& Gupta, G. P.} {1999}]{1999Tayal_apj526} Tayal, S. S. \& Gupta, G. P. , 1999, ApJ, 526,  544-548

\bibitem[\protect\citeauthoryear{Tayal, S. S. \& Zatsarinny, O.} {2010}]{2010Tayal_apjs188} Tayal, S. S. \& Zatsarinny, O. , 2010, ApJ, 188,  32-45

\bibitem[\protect\citeauthoryear{Tayal, S. S.} {2011}]{2011Tayal_apjs195} Tayal, S. S. , 2011, ApJ, 195,  12

\bibitem[\protect\citeauthoryear{Todt et al.}{2015}]{Todt2015} Todt H., Sander A., Hainich R., Hamann W.-R., Quade M., Shenar T., 2015, A\&A, 579, A75. 


\bibitem[{{Todt} {et~al.}(2008){Todt}, {Hamann}, \& {Gr{\"a}fener}}]{Todt2008} {Todt}, H., {Hamann}, W.-R., \& {Gr{\"a}fener}, G. 2008, in Clumping in  Hot-Star Winds, ed. W.-R. {Hamann}, A.~{Feldmeier}, \& L.~M. {Oskinova}, 251

\bibitem[\protect\citeauthoryear{Weidmann, et al.}{2020}]{Weidmann2020} Weidmann W.~A., et al., 2020, arXiv, arXiv:2005.10368


\bibitem[Werner et al.(2007)]{Werner2007} Werner, K., Rauch, T., \& Kruk, J.~W.\ 2007, \aap, 474, 591




\bibitem[\protect\citeauthoryear{Zeippen, C. J.} {1982}]{1982Zeippen_mnra198} Zeippen, C. J. , 1982, MNRAS, 198,  111-125



\end{thebibliography}
